\documentclass[aps,preprint,preprintnumbers,showpacs,onecolumn]{revtex4}

\renewcommand{\vec}[1]{\mathbf{#1}}

\newcommand{\SP}{\textit{single particle}}
\newcommand{\CP}{\textit{cluster particle}}

\newcommand{\SPS}{\textit{single particles}}
\newcommand{\CPS}{\textit{cluster particles}}
\newcommand{\CSPS}{\textit{Single particles}}

\newcommand{\B}{\Gamma}

\usepackage{graphics}
\usepackage{graphicx}
\usepackage{amssymb,bm,amsmath}
\usepackage{color}
\usepackage{verbatim}
\usepackage{float}
\newcommand{\grad}{\bm{\partial}}
\newcommand{\R}{\mathfrak{Re}}
\newcommand{\I}{\mathfrak{Im}}

\begin{document}

\title[Role of particle conservation in self-propelled particle systems]{Role of particle conservation in self-propelled particle systems}

\author{Christoph A.Weber$^{1,2}$,  Florian Th\"uroff$^{1,2}$  and Erwin Frey$^1$}

\address{$^1$Arnold Sommerfeld Center for Theoretical Physics and Center for NanoScience, Department of Physics, Ludwig-Maximilians-Universit\"at M\"unchen, Theresienstr. 37, D--80333 Munich, Germany}
\address{$^2$C.A. Weber and F. Th\"uroff contributed equally to this work.}
\preprint{LMU-ASC 07/13, NSF-KITP-12-120}

\email{frey@lmu.de}

\begin{abstract}
Actively propelled particles undergoing dissipative collisions are known to
develop a state of spatially distributed coherently moving clusters.  For densities larger than a
characteristic value clusters grow in time and form a stationary well-ordered state of coherent macroscopic motion.
In this work we address two questions: 
(i) What is the role of the particles' aspect ratio in the context of cluster formation, and does the particle shape affect the system's behavior on hydrodynamic scales?
(ii) To what extent does particle conservation influence pattern formation?
To answer these questions we suggest a simple kinetic model permitting to depict some of the interaction properties between freely moving particles and particles integrated in clusters. To this end,  we introduce two particle species:  single and cluster particles. 
 Specifically, we account for coalescence of clusters from single particles, assembly of single particles on existing clusters, collisions between clusters, and cluster disassembly.
 Coarse-graining our kinetic model, (i) we demonstrate that particle shape (i.e. aspect ratio) shifts the scale of the transition density, but does not impact
the instabilities at the ordering threshold.
(ii) We show that the validity of particle conservation determines the existence of a longitudinal instability, which tends to amplify density heterogeneities locally, and in turn triggers a wave pattern with wave vectors parallel to the axis of macroscopic order. 
If the system is in contact with a particle reservoir
this instability vanishes
due to a compensation of 
density heterogeneities. 
\end{abstract}



\maketitle

\section{Introduction}
\label{intro}

The emergence of collective motion is a ubiquitous phenomenon in nature, encountered in a great variety of actively propelled systems \cite{Ramaswamy_Review,Aranson_Tsimring_book,Marchetti:2012ws}. Coherently moving groups have been observed over a broad range of length scales, spanning from micrometer-sized systems \cite{Butt,Schaller,schaller2,pnas_bacteria,Dombrowski_2004,Ringe_PNAS,Yutaka}   over millimeter large granules \cite{Dachot_Chate_2010,Dauchot_long,Kudrolli_2008}  to large groups of animals \cite{ballerini_starlings}. 
The fact that the capability of synchronizing movements between agents is shared even among fundamentally different systems has called for abstract modeling approaches, aiming at identifying the essential properties of these systems both, in terms of analytical descriptions \cite{Toner_Tu_1995,Aronson_MT,Bertin_short,Aranson_Bakterien,Marchetti2008,Baskaran_Marchetti_2008,Holm_Putkaradze_Tronci_2008,Bertin_long,Holm_Putkaradze_Tronci_2010,Mishra_Marchetti_2010,Ihle_2011,Peshkov:2012uu,Peshkov:2012tu,Wensink04092012}, and by means of agent based simulation techniques \cite{Vicsek,Czirok_2000,Gregoire:2004uf,Chate_long,Chate_Variations,Grossman_Aranson,Albano2009,Ginelli,Peruani_rods,Peruani_Traffic_Jams,Weber_nucleation}.

Theoretically, the emergence of collective motion has mostly been studied in the context of particle conserving systems. There are, however, a number of experimental systems, in which the assumption of particle conservation is questionable. In typical gliding assays \cite{Butt,Schaller,schaller2,Ringe_PNAS,Yutaka}, for instance, collective motion of filaments is observed on a two-dimensional ``motor carpet'' which itself is in contact with a three-dimensional bulk reservoir of filaments. However, the impact of particle conservation on the formation of patterns of collective motion remains largely elusive.

Here, we address the significance of constraints for particle number by highlighting the differences in the collective properties between particle conserving systems and those in contact with a particle reservoir. Our focus will be on the comparison of two archetypical scenarios, which we will refer to as the \emph{canonical} (particle conserving), and the \emph{grand canonical} (violating particle conservation) scenario, respectively.

To this end, we will resort to a kinetic approach, which has been set up previously by Aranson et al. \cite{Aronson_MT} to describe pattern formation in a system of interacting microtubules, and which has been extended to the case of self-propelled spheres by Bertin et al. \cite{Bertin_short,Bertin_long}. In the following we will extend this description 
in accordance with a physical picture of collective motion  
that has been developed over the last decade based on observations in agent-based simulations of locally interacting, particle conserving systems   \cite{Gregoire:2004uf,Chate_long,Peruani_rods,Weber_nucleation}. Among the most pertinent phenomena that have been reported in the context of these studies is the formation of intricate local structures pervading these systems in the vicinity of the ordering transition: Densely packed cohorts of coherently moving particles---subsequently referred to as \emph{clusters}---incessantly ``nucleate'' and ``evaporate'' on local scales, even below threshold, rendering the system isotropic and homogeneous only in the limit of macroscopic length scales. Individual particles exhibit superdiffusive behavior in this regime, performing quasi-ballistic ``flights'' as long as they are part of a cluster, and conventional particle diffusion if they are not. Above threshold, collective motion manifests itself on macroscopic scales in the form of coherently moving and dense bands, which are submersed in an isotropic low-density ``particle sea''. Spatially homogeneous flowing states, in contrast, are observed only well beyond the ordering threshold \cite{Gregoire:2004uf}.
Moreover, particle geometry was demonstrated to play an essential role in the context of clustering dynamics, with higher aspect ratios facilitating the formation of clusters of coherently moving particles \cite{Peruani_rods}.

\begin{figure}
	\centering
	\resizebox{0.9\textwidth}{!}{%
	 \includegraphics{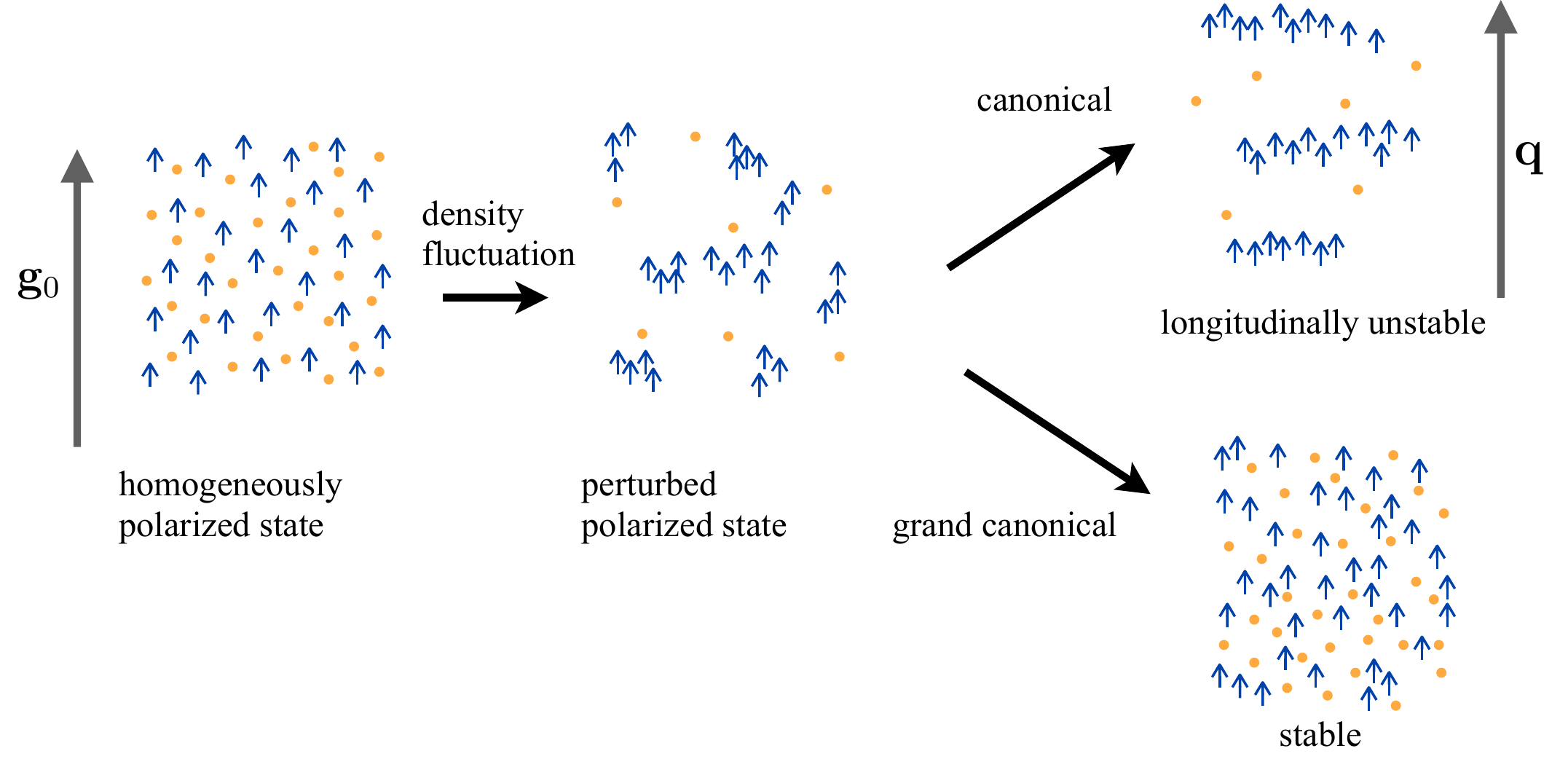}}
	 	\caption{Illustration of the \emph{canonical} and \emph{grand canonical} modeling framework, highlighting the quintessential differences in the context of pattern formation. In the homogeneously polarized state (left), the \CPS~density (blue arrows) constitutes the system's macroscopic net momentum $\vec g_0$, while some fraction of the system's particles, the  \SPS~(orange dots) exhibit zero net momentum. Spatial perturbations of both density fields lead to two fundamentally different outcomes: (i) In case of a closed system obeying total particle conservation (\SPS $+$\CPS), termed as the \emph{canonical} model, the homogeneously polarized state is longitudinally unstable, with a wave vector $\vec q$ parallel to the polarized state $\vec g_0$, potentially enforcing a wave-like pattern. (ii) In contrast, open systems turn out to be stable against this kind of density fluctuations.
	}
	\label{<fig:IlluMotivation>}
\end{figure}

In the light of the above, we suggest a simplified modeling framework  
 to incorporate the intricate role of clusters on the ordering behavior, which will be presented in greater technical detail in the following section: Particles interact via binary collisions with a scattering cross section which is explicitly derived as a function of particle shape. Depending on whether a given particle is part of a cluster or not, it will be associated with one of two distinct particle classes, which we will refer to as the class of \CPS~and the class of \SPS, respectively. \CSPS~are ``converted'' to \CPS~by ``condensation'' every time a \SP~collides with a cluster. Conversely, \CPS~are ``converted'' back to \SPS~by an ``evaporation'' process which we assume to occur at some constant (possibly particle shape dependent \cite{Peruani_rods}) rate. Moreover, in the absence of interactions, \CPS~will be assumed to move ballistically, whereas \SPS~will be assumed to perform random walks. Taken together, the conversion dynamics and the class-specificity of particle motion provide a simple way to implement the typical superdiffusive behavior of individual particles, that was alluded to above. To assess the importance of particle conservation in the context of pattern formation, we will analyze two variants of this model: Firstly, we study closed systems in which the total number of particles is conserved (\emph{canonical} scenario) and where, consequently, the denser cluster phase grows at the expense of the single phase. Secondly, we examine open systems in contact with a particle reservoir (\emph{grand canonical} scenario), where the particle current out of the single phase is compensated so as to retain the density of the isotropic sea of \SPS~at a constant level; cf. figure \ref{<fig:IlluMotivation>}.
 
Our work is structured as follows: 
In \ref{<sect:model>} the modeling framework for the \emph{canonical} and \emph{grand canonical} model is introduced and the model equations are discussed in detail. The corresponding hydrodynamic equations are derived in \ref{<sect:hydro_dev>} by means of an appropriate truncation scheme in Fourier space. Therein, we also give explicit expressions of the kinetic coefficients as a function of the particles' aspect ratio and velocity, noise level and density for \SPS~and \CPS.  
\ref{<sect:homogeneous>} is devoted to the analysis of the homogeneous equations. The dynamic's stationary fixed points are determined and the phase boundary between the isotropic and homogeneous state is calculated. 
\ref{<sect:inhomogeneous>} deals with the implications of the inhomogeneous equations in the framework of a linear stability analysis, which are concluded in \ref{<sect:conclusion>}. 



\section{Coarse-grained kinetic model}\label{<sect:model>}

We consider rod-like particles of length $L$ and diameter $d$ moving in two dimensions with a constant velocity $v$. A particle's state is determined by its position $\vec x$ and the orientation $\theta$ of its velocity vector. To describe the time evolution of the system, we adopt a kinetic approach~\cite{Aronson_MT,Bertin_short,Aranson_Bakterien,Bertin_long}. 

On mesoscopic scales, the system's spatio-temporal evolution is then governed by Boltzmann-like equations for the one-particle distribution functions within the classes of \SPS~and \CPS, respectively.
Interactions enter this description by means of collision integrals. The kernel of these integrals involves both, a measure for the rate of collisions, as well as a ``collision rule''  implementing a mapping between pre- and post-collisional directions $\theta$ and $\theta'$ of each of the two partaking particles.
Here, we are led to consider a simplified model of binary particle interactions, which builds on the distinction between \SPS~and \CPS. The details of this model will be described in the following section.

 \subsection{\label{sec:ReactionEquations}Reaction equations}
  
Let $S(\theta)$ and $C(\theta)$ refer to a particle moving in the direction of $\theta$ and being associated with the class of \SPS~or \CPS, respectively. In the absence of interactions, \SPS~are assumed to perform a persistent random walk, which we model as a succession of ballistic straight flights, interspersed by self-diffusion (``tumble'') events. These tumble events are assumed to occur at a constant rate $\lambda$ and reorient the particle's orientation $\theta$ by a random amount $\vartheta_0$:
 \begin{equation}\label{<StoS>}
	S(\theta)  \overset{\lambda}{\rightarrow}	S(\theta'= \theta + \vartheta_{0}). 
\end{equation} 
   For simplicity we assume $\vartheta_0$ to be Gaussian-distributed,
 \begin{equation}\label{<Gauss_singles>}
	p_{0}(\vartheta_0) = \frac{1}{\sqrt{2 \pi \sigma_0^2}}  \exp{\big(-\vartheta_0^2/2 \sigma_{0}^2\big)},
\end{equation}
with $\sigma_0$ denoting the standard deviation. On times scales much larger than $\lambda^{-1}$, this tumbling behavior can be described as conventional particle diffusion, with the particles' diffusion constant being a function of  $\lambda$ and $\sigma_0$ \cite{Lovely1975477}.

\begin{figure}
	\centering
	\resizebox{0.48\textwidth}{!}{%
	 \includegraphics{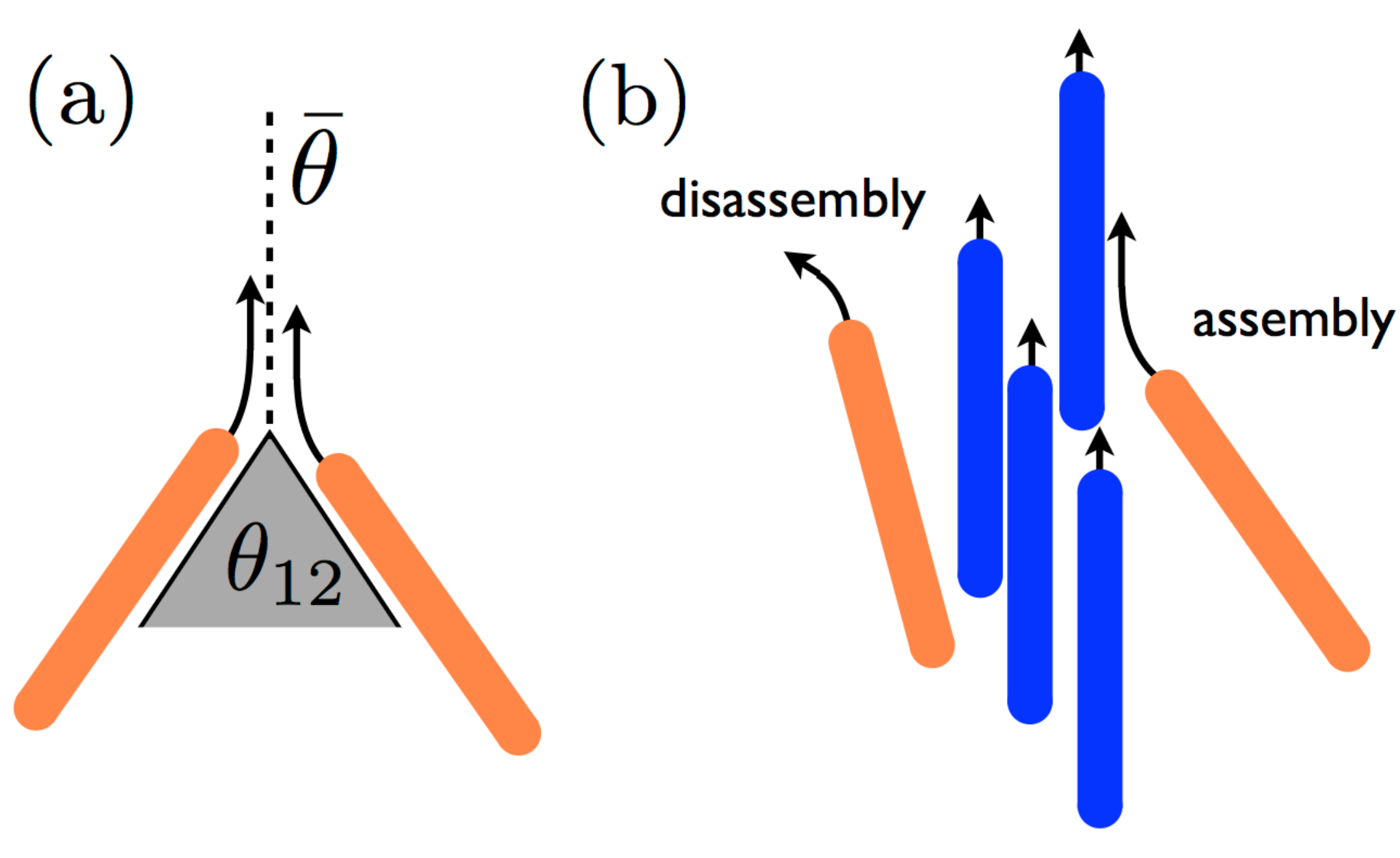}}
			\caption{(a) Illustration of two single particle species (light orange) with a pre-collisional relative angle of $\theta_{12}$, colliding such that they align collinear to the average angle $\bar \theta$. Both particles become a cluster species after the collision.  (b) Right: Illustration of a possible scenario, where a single particle joins a cluster by perfectly aligning to the cluster particles (blue). Left: A particle leaves the cluster by a random change of its direction at a characteristic rate $\epsilon$. 
	}
	\label{fig:illustration}
\end{figure}

When two \SPS~$S(\theta_1)$ and $S(\theta_2)$ collide they are assumed to assemble a cluster, i.e. each of the two particles becomes a \CP~(see  figure \ref{fig:illustration}a):   
\begin{equation}\label{<SStoCC>}
	S(\theta_1) + S(\theta_2)  \rightarrow  C(\bar \theta + \vartheta) + C(\bar \theta + \vartheta),
\end{equation} 
where \footnote{To make sure that $\bar\theta$ points into the ``right'' direction (i.e. $|\bar\theta-\theta_{1/2}|\leq\pi/2$), we choose $\theta_1\in(-\pi,\pi]$ and $\theta_2\in(\theta_1-\pi,\theta_1+\pi]$.}
\begin{equation}\label{<polar_rule>}
	\bar \theta(\theta_1, \theta_2) = \frac{1}{2} (\theta_1 + \theta_2)
\end{equation}
denotes the average of both pre-collisional angles $\theta_1$ and $\theta_2$, and where $\vartheta$ is a random variable which we, again, assume to be Gaussian-distributed:
\begin{equation}\label{<Gauss_cluster>}
	p(\vartheta) = \frac{1}{\sqrt{2 \pi \sigma^2}}  \exp{\big(-\vartheta^2/2 \sigma^2\big)}.
\end{equation}
The rate of binary collisions, such as equation \eqref{<SStoCC>}, are determined by a particle-shape dependent \emph{differential scattering cross section}, which will be discussed below; see \ref{<sect:constitutive_eqs>} and  \ref{<app:stosszylinder>}.

 Collisions involving \CPS~are distinct from \SP~events. Due to the close spatial proximity of particles within each cluster these collisions correspond to many-particle interactions. Needless to say, a detailed description of cluster formation and the ensuing particle dynamics represents a highly complex matter, requiring explicit consideration of such many-particle interactions. For simplicity, we will resort to the following simplified interaction picture: We assume that (binary) collisions between \SPS~and \CPS~lead to a condensation process during which the \SP~aligns to the \CP~without changing the direction of the cluster as a whole:
\begin{equation}\label{<SCtoCC>}
	S(\theta_1) + C(\theta_2) \rightarrow  C(\theta_2) + C(\theta_2).
\end{equation} 
Eq. \eqref{<SCtoCC>} thus captures the net effect of collisions between \SPS~and \CPS, during which multiple collisions, involving neighboring particles belonging to the same cluster, stabilize the cluster's direction; cf. figure \ref{fig:illustration} b for an illustration. 

Collisions among \CPS~is an even more intricate process, since they actually depend on size and shape of both colliding clusters, and in general involve multi-particle interactions.
In the framework of a Boltzmann-like description, correlations in the particle distribution are neglected and only binary interactions are considered. The frequency of interactions are determined by a geometrical construction called the ``Boltzmann cylinder'', assuming that particle positions are homogeneously distributed on local scales.
With regard to many-particle interactions during collisions among \CPS, we thus have to resort to some kind of simplified, binary collision picture.
Since our kinetic model lacks any direct notion of cluster size or shape, we will stick to the assumption that, on average, collisions between \CPS~are devoid of any directional bias, leading to the same type of collision rule as for \SPS:
\begin{equation}\label{<CCtoCC>}
	C(\theta_1) + C(\theta_2) \rightarrow  C(\bar \theta + \vartheta) + C(\bar \theta + \vartheta).
\end{equation}
Again, $\vartheta$ constitutes a Gaussian-distributed random variable  given in \eqref{<Gauss_cluster>}.
Moreover, due to external (e.g. thermal background) and internal (e.g. noisy propelling mechanism) noise, \CPS~evaporate to become \SPS.
In analogy to the self-diffusion of \SPS, we thus introduce a rate \footnote{As has been pointed out in Ref. \cite{Peruani_rods}, this rate may depend on particle shape.} $\epsilon$ characterizing the following evaporation process:
\begin{equation}\label{<CtoS>}
	C(\theta)  \overset{\epsilon}{\rightarrow}  S(\theta' =\theta + \vartheta_0). 
\end{equation} 
Also in this case, the strength of the angular changes are Gaussian-distributed according to \eqref{<Gauss_singles>}, and for simplicity we use the same standard deviation $\sigma_0$ as for the \SPS' persistent random walk. As discussed above, \CPS~are strongly caged due to their close proximity to neighboring, collinearly moving particles. Reorientations of \CPS~due to noise are therefore strongly counteracted by realigning particle collisions, rendering \CPS~considerably less susceptible to random fluctuations than \SPS. Hence, we assume 
\begin{equation}
\label{<eq:lambdaGGepsilon>}
\epsilon\ll\lambda,
\end{equation}
which is consistent with the observations in agent-based simulations slightly below  the ordering transition  \cite{Gregoire:2004uf}, finding coherently moving clusters in an unpolarized background of randomly moving particles. In this regime individual particles exhibit superdiffusive behavior, performing quasi-ballistic ``flights'' as long as they are part of a cluster, and conventional particle diffusion if they are not.

\subsection{Constitutive equations}
\label{<sect:constitutive_eqs>}

Building on the modeling framework defined above, we now set up a kinetic description for the \emph{canonical} model. We denote by $s(\theta,\vec{x},t)$ and $c(\theta,\vec{x},t)$ the one-particle distribution functions within the class of \SPS~and \CPS, respectively, i.e. $s(\theta,\vec{x},t)\,d\theta\,d^2x$ gives the number of \SPS~located in an infinitesimal region $[\vec x,\vec x+d\vec x]$ with orientations in the interval $[\theta,\theta+d\theta]$ (and likewise for $c(\theta,\vec{x},t)\,d\theta\,d^2x$). Both one-particle distribution functions are subject to convection due to the propelling velocity $\vec v$ of each particle. Moreover, local fluctuations in the one-particle distribution functions due to self-diffusion and collision events are to be accounted for. We thus arrive at the following set of Boltzmann-like equations for the \emph{canonical} model:
\begin{subequations}
\begin{eqnarray}
\label{<eq:BoltzmannEqsS>}
\partial_t s(\theta,\vec{x},t)+\vec{v}\cdot\nabla s(\theta,\vec{x},t) &=& \dot s(\theta,\vec{x},t),\\
\label{<eq:BoltzmannEqsC>}
\partial_t c(\theta,\vec{x},t)+\vec{v}\cdot\nabla c(\theta,\vec{x},t) &=& \dot c(\theta,\vec{x},t),
\end{eqnarray}
\end{subequations}
where the source terms $\dot s(\theta,\vec{x},t)$ and $\dot c(\theta,\vec{x},t)$ read 
\begin{subequations}
\label{eq:Source}
\begin{eqnarray}
\label{eq:SourceS}
\dot s &=& \lambda\left[\mathcal{D}^{(+)}_s(\theta)-\mathcal{D}^{(-)}_s(\theta)\right]+\epsilon\mathcal{D}^{(+)}_c(\theta) -\mathcal{C}^{(-)}_s(\theta)-\mathcal{A}[s,c;\theta],\\
\label{eq:SourceC}
\dot c &=& -\epsilon \mathcal{D}^{(-)}_c(\theta)+\mathcal{C}^{(+)}_s(\theta)+\mathcal{C}^{(+)}_c(\theta)+\mathcal{A}[c,s;\theta]-\mathcal{C}^{(-)}_{c}(\theta) \, .
\end{eqnarray}
\end{subequations}
They give the net number of \SPS~and \CPS~entering the phase space region $d\omega=[\vec x,\vec x+d\vec x]\times[\theta,\theta+d\theta]$ per unit time and unit area, respectively. The various terms correspond to gain [superscript$^{(+)}$] and loss [superscript$^{(-)}$] of particles by the following processes: 


(i) \emph{Self-diffusion and evaporation}. In these cases the source terms are products of the corresponding rates and probability densities, with
\begin{eqnarray}
\mathcal{D}^{(-)}_{f}(\theta) &=& f(\theta)
\end{eqnarray}
denoting the probability density for a particular species $f$ to have a certain angle $\theta$, and 
\begin{eqnarray}
\mathcal{D}^{(+)}_{f}(\theta) &=& \left \langle f(\theta-\vartheta_0) \right\rangle_0
\end{eqnarray}
denoting the transition probability from $\theta' = \theta - \vartheta_0$ to $\theta$ averaged over all $\vartheta_0$ with respect to the Gaussian weight \eqref{<Gauss_singles>}. Note that, here and in the following, the argument of $f$ is understood modulo $2 \pi$.

(ii) \emph{Collisions within the same class of particles}. The collision integrals, representing the processes defined in equations \eqref{<SStoCC>} and \eqref{<CCtoCC>}, are given by standard expressions \cite{Aronson_MT,Bertin_short,Aranson_Bakterien,Bertin_long}
\begin{subequations}
\begin{eqnarray}
\mathcal{C}^{(+)}_f(\theta) &=& \left\langle \int d\mathcal{I}\,f(\theta')f(\theta'') \, \delta\bigg(\bar \theta(\theta', \theta'')+\vartheta-\theta\bigg) \right\rangle, \\
\mathcal{C}^{(-)}_f(\theta) &=& \int d\mathcal{I}\,f(\theta')f(\theta'')\delta(\theta'-\theta).
\end{eqnarray}
\end{subequations}
Here $\langle ... \rangle$ denotes an average over $\vartheta\in(-\infty, \infty)$ with respect to the Gaussian weight \eqref{<Gauss_cluster>} and the average angle $\bar \theta$ is given in Eq. \eqref{<polar_rule>}. The integral measure 
\begin{equation}
\int d\mathcal{I}\, (\ldots) \equiv \int_{-\pi}^{\pi}d\theta' \int_{\theta'-\pi}^{\theta'+\pi}d\theta'' \, \B(L,d,|\theta'-\theta''|)\,(\ldots) \, ,
\end{equation}
contains the \emph{differential scattering cross section}
\begin{equation}
\B(L,d,|\theta'-\theta''|)=4dv\left|\sin\left(\frac{\theta'-\theta''}{2}\right)\right|\left[1+\frac{(L/d)-1}{2}\left|\sin(\theta'-\theta'')\right|\right]
\end{equation}
characterizing the frequency of collisions (i.e. hard-core interactions) between rod-like particles. The scattering function $\B$ itself carries all information concerning the shape of the particles and is a function of the relative orientation of the colliding particles.
Reminiscent of the Boltzmann scattering cylinder, $\B$ can be derived on the basis of purely geometric considerations  assuming that all spatial coordinates within the cylinder are equally probable; for details see \ref{<app:stosszylinder>}. 

(iii) \emph{Assembly events of a \SP~joining a cluster}. These events, represented by \eqref{<SCtoCC>}, occur through binary collisions between \SPS~and \CPS~and are thus represented by an analogous integral expression:
\begin{eqnarray}
\mathcal{A}[f,g;\theta] &=& \int_{}^{}d\mathcal{I}\,f(\theta')g(\theta'')\delta(\theta'-\theta).
\end{eqnarray}


\section{Derivation of hydrodynamic equations}\label{<sect:hydro_dev>}

In order to reduce our kinetic description to a set of hydrodynamic equations valid on large length and time scales, we follow the well-established procedure of Aranson et al. \cite{Aronson_MT} and Bertin et al. \cite{Bertin_short,Bertin_long}, and analyze the angular dependence of equations \eqref{<eq:BoltzmannEqsS>} and  \eqref{<eq:BoltzmannEqsC>} in Fourier space. Due to the $2\pi$-periodicity in $\theta$, the one particle distribution functions can be expanded in Fourier series
\begin{subequations}
\begin{eqnarray}
\label{<eq:FourierS>}
s(\theta, \vec x,t) &=& \frac{1}{2\pi} \sum_{n=-\infty}^{\infty} s_n(\vec x,t) e^{-in\theta},\\
\label{<eq:FourierC>}
c(\theta, \vec x,t) &=& \frac{1}{2\pi}\sum_{n=-\infty}^{\infty} c_n(\vec x,t) e^{-in\theta},
\end{eqnarray}
\end{subequations}
where
\begin{subequations}
\begin{eqnarray}
\label{<eq:FTs>}
s_n(\vec x,t) &=& \int_{-\pi}^{\pi}d\theta\,e^{i n \theta}s(\theta,\vec x,t),\\
\label{<eq:FTc>}
c_n(\vec x,t) &=& \int_{-\pi}^{\pi}d\theta\,e^{i n \theta}c(\theta,\vec x,t).
\end{eqnarray}
\end{subequations}
Upon identifying 
$\mathbb R^2\leftrightarrow\mathbb C$, e.g.  $\vec v\leftrightarrow v\,e^{i\theta}$ ($v=|\vec v|$), the zeroth and first Fourier modes are directly connected to the hydrodynamic densities $\rho_s$ (\SP~density) and $\rho_c$ (\CP~density), and the corresponding current density $\vec g_s$ and $\vec g_c$ , i.e.
\begin{subequations}
\begin{eqnarray}
\label{<eq:HydrodynamicMomentsS>}
\rho_s(\vec x,t) &=& s_0(\vec x,t),\\
\label{<eq:HydrodynamicMomentsC>}
\rho_c(\vec x,t) &=& c_0(\vec x,t),\\
\label{<eq:HydrodynamicMomentsGS>}
\vec g_s(\vec x,t)\equiv\rho_s(\vec x,t)\,\vec u_s(\vec x,t) &=& v\,s_1(\vec x,t),\\
\label{<eq:HydrodynamicMomentsGC>}
\vec g_c(\vec x,t)\equiv\rho_c(\vec x,t)\,\vec u_c(\vec x,t) &=& v\,c_1(\vec x,t).
\end{eqnarray}
\end{subequations}
In equations \eqref{<eq:HydrodynamicMomentsGS>} and \eqref{<eq:HydrodynamicMomentsGC>}, the ``='' signs indicate identification of vectors and complex numbers. The quantities $\vec u_{s/c}$ denote the velocities of the macroscopic flow fields established by \SPS~and \CPS, respectively. Also note that the second Fourier components are proportional to the nematic order parameter within the respective class of particles (as reflected by the symmetry of $e^{i2\theta}$ under $\theta\rightarrow\theta+\pi$).
Using equations \eqref{<eq:FourierS>} and \eqref{<eq:FourierC>}, the Boltzmann-like equations \eqref{<eq:BoltzmannEqsS>} and  \eqref{<eq:BoltzmannEqsC>} transform to
\begin{subequations}
\begin{eqnarray}
\label{<eq:Fourier_S>}
&&\partial_ts_k+\frac{v}{2}\bigg[\partial _x\left(s_{k+1}+s_{k-1}\right)-i\partial _y\left(s_{k+1}-s_{k-1}\right)\bigg]=\\
\nonumber
&&-\lambda s_k+e^{(k\sigma_0)^2/2}\left(\lambda s_k+\epsilon c_k\right)-\sum_{n=-\infty}^{\infty}I_{n,0}\left(s_n+c_n\right)s_{k-n} \,\\
\label{<eq:Fourier_C>}
&&\partial_tc_k+\frac{v}{2}\bigg[\partial _x\left(c_{k+1}+c_{k-1}\right)-i\partial _y\left(c_{k+1}-c_{k-1}\right)\bigg]=\\
\nonumber
&&-\epsilon c_k + \sum_{n=-\infty}^{\infty}\left[I_{n,0}(s_nc_{k-n}-c_nc_{k-n})+e^{(k\sigma)^2/2}I_{n,k}\left(s_ns_{k-n}+c_nc_{k-n}\right)\right],
\end{eqnarray}
\end{subequations}
where the collision integrals $I_{n,k}$ are defined as follows:
\begin{equation}
\label{<eq:Ink>}
I_{n,k}=\frac{1}{2\pi}\int_{-\pi}^{\pi}d\phi\,\B(L,d,|\phi|)\cos\left[\left(n-\frac{k}{2}\right)\phi\right].
\end{equation}
Note, in particular, that $I_{0,0}$ gives the \emph{total scattering cross section}.

\subsection{Truncation scheme}\label{<sect:truncation>}

 Equations \eqref{<eq:Fourier_S>} and \eqref{<eq:Fourier_C>} constitute an infinite set of coupled equations in Fourier space, which are fully equivalent to the Boltzmann-like equations \eqref{<eq:BoltzmannEqsS>} and \eqref{<eq:BoltzmannEqsC>}. To derive a closed set of hydrodynamic equations, we need to consider some additional assumptions, allowing us to truncate this infinite Fourier space representation.
 
Here, our focus will be on virtually isotropic systems in the vicinity of an ordering transition breaking rotational symmetry. In this case, deviations of the one-particle distribution functions from the constant distribution $\sim 1/2\pi$ are small and contributions from large wave numbers in the Fourier series \eqref{<eq:Fourier_S>} and \eqref{<eq:Fourier_C>} are negligible. 
We further consider sufficiently dilute systems, in which the number of (binary) particle collisions per unit time and area [$\sim(\rho_c+\rho_s)^2\,I_{0,0}$] is much smaller than the corresponding number of \SP~diffusion events [$\sim\lambda\,\rho_s$]. 
Together with $\epsilon \ll \lambda$ [equation \eqref{<eq:lambdaGGepsilon>}], stating that disassembly from a cluster is strongly hindered by particle caging, allows us to treat \SP~diffusion as a fast process. The \SP~phase thus acts as an isotropic sea of particles where particle orientations (but not necessarily particle densities) are equilibrated, and hence the net hydrodynamic flow vanishes [$\vec u_s=0$].
Finally, from a dimensional analysis of equations \eqref{<eq:FTs>} and \eqref{<eq:FTc>}, together with \eqref{<eq:HydrodynamicMomentsGS>} and \eqref{<eq:HydrodynamicMomentsGC>}, one finds $c_k/\rho_c \sim \mathcal{O}(|\vec u_c|^k/v^k)$. Near the onset of order, where $|\vec u_c|/v\ll1$, we only consider the density ($c_0$) and polarity ($c_1$) of \CPS, and use the stationary equation for $c_2$ as a closure relation, neglecting all contributions from higher order coefficients.

In summary, we resort to the following truncation scheme, leading to a set of hydrodynamic equations, valid near the onset of the ordering transition:
\begin{subequations}
\label{<eq:truncation>}
\begin{eqnarray}
\label{<eq:s_truncation>}
s_k &=& 0, \qquad\forall |k|>0,\\
\label{<eq:c_truncation>}
c_k &=& 0, \qquad\forall |k|>2.
\end{eqnarray}
\end{subequations}

\subsection{Derivation of the hydrodynamic equations}

With the above truncation scheme, \eqref{<eq:Fourier_S>} and \eqref{<eq:Fourier_C>} reduce to
\begin{subequations}
\begin{eqnarray}
\label{<eq:dts0>}
\partial_ts_0 &=& \epsilon c_0-I_{0,0}\left(s_0^2+s_0c_0\right),\\
\label{<eq:dtc0>}
\partial_tc_0 &=& -v\left[\partial_x \R(c_1)+\partial_y\I(c_1)\right]-\partial_ts_0,\\
\label{<eq:dtc1>}
\partial_tc_1 &=& -\frac{v}{2}\left[\partial_x(c_2+c_0)-i\partial_y(c_2-c_0)\right]\\
\nonumber
&&+ \left[  \left(2e^{-\sigma^2/2} I_{1,1} - I_{1,0} - I_{0,0}\right)  c_0 -\epsilon + I_{0,0} s_0\right] c_1 \\
\nonumber
&&+ \left[  2e^{-\sigma^2/2} I_{2,1} - I_{1,0} - I_{2,0} \right] c_1^* c_2, \\
\label{<eq:dtc2>}
\partial_t c_2 &=& -\frac{v}{2}\left[\partial_x+i\partial_y\right]c_1\\
\nonumber
&&+ \left[  \left(2e^{-2\sigma^2} I_{1,0} - I_{2,0} - I_{0,0} \right) c_0 -\epsilon + I_{0,0} s_0 \right] c_2 \\
\nonumber
&&+ \left[  e^{-2\sigma^2} I_{0,0} - I_{1,0}  \right] c_1 c_1,  
\end{eqnarray}
\end{subequations}
where we used $f_{-k}=f_k^*$, since $f(\theta)\in\mathbb R$ ($f\in\{s,c\}$), and where $\R(a)$ [$\I(a)$] denote the real [imaginary] part of $a$. Moreover, as can be seen from the definition in \eqref{<eq:Ink>}, the collision integrals $I_{n,k}$ only depend on the value $|n-k/2|$, whence only five of the collision integrals appearing in the above equations are independent. These integrals as a function of the particle's aspect ratio are evaluated and summarized in table \ref{<tab:values_of_Ink>}. Also note that the entire set of equations \eqref{<eq:dts0>} -- \eqref{<eq:dtc2>} is independent of the fast \SP~diffusion time scale $\lambda^{-1}$ (and, hence, also of the diffusion noise parameter $\sigma_0$). In our present approach, $\lambda$ has only a conceptual meaning in maintaining a well-mixed particle bath within the class of \SPS.

For given particle densities, the time scales governing the dynamics of the polar and nematic order parameter fields, represented by $c_1$ and $c_2$, are given by the linear coefficients in the second line of  \eqref{<eq:dtc1>} and \eqref{<eq:dtc2>}, respectively. 
As will be detailed in  \ref{<sect:homo_pol>}, the onset of collective motion is hallmarked by a change in sign of the linear coefficient in \eqref{<eq:dtc1>}, implying a diverging time scale for the dynamics of the polarity field. 
On the other hand the time scale for $c_2$ is finite for all densities
which implies that the relaxation of the nematic order parameter field is fast compared to the polarity field. This allows us to set 
$\partial_tc_2\approx0$ in  \eqref{<eq:dtc2>}.

In the following it will be convenient to write down equations in dimensionless form. To this end, we construct the following characteristic scales: Time and space will be measured in units of the cluster evaporation time and length scale 
\begin{equation}\label{<tau_e>}
\hat \tau_e=\epsilon^{-1} \quad\mbox{and}\quad \hat\ell_e=v/\epsilon.
\end{equation}
From the cluster evaporation time scale $\hat{\tau}_e$ and the \emph{total scattering cross section} $I_{0,0}$, we can construct the characteristic density scale
\begin{equation}\label{<rho_b>}
\hat\rho_b = \frac{1}{I_{0,0}\,\hat\tau_e}.
\end{equation}
The \SP~and \CP~phase constantly exchange particles at rates that are determined by cluster evaporation ($\epsilon$) on the one hand (\CPS~$\rightarrow$ \SPS), and cluster nucleation due to particle collisions on the other hand (\SPS~$\rightarrow$ \CPS), which occur with a rate $\sim\rho\,I_{0,0}$.
Therefore the characteristic density scale $\hat\rho_b$ marks the particle density, where both rates balance. 
In particular, $\rho / \hat{\rho}_b=(\rho_s+\rho_c)/ \hat{\rho}_b$ gives the rate of inter-particle collisions relative to cluster evaporation events. Thus, the numerical quantity $\rho/ \hat{\rho}_b$ provides a direct measure expressing the competition between the randomizing effects of noise and the order creating effects of particle collisions, hallmarking the onset (and maintenance) of collective motion \cite{Vicsek}.

We thus arrive at the following rescaling scheme
\begin{subequations}
\begin{eqnarray}
\label{<equ:dens_scale>}
t &\rightarrow& t\cdot\hat\tau_e,\\
\vec{x} &\rightarrow& \vec{x}\cdot\hat\ell_e,\\
\rho_{s/c} &\rightarrow& \rho_{s/c}\cdot\hat\rho_b,\\
\vec{g} &\rightarrow& \vec{g}\cdot \hat\rho_b\,\frac{\hat\ell_e}{\hat\tau_e},\\
\label{<equ:dens_scale5>}
I_{n,k} &\rightarrow& I_{n,k}\cdot \frac{1}{\hat\rho_b\,\hat\tau_e},
\end{eqnarray}
\end{subequations}
where the characteristic scales for momentum ($\vec{g}$) and scattering cross section ($I_{n,k}$) have been constructed from those of time, space, and density. In this rescaling the momentum current density is equal to one if the corresponding fluid element with a characteristic density $\hat\rho_b$, for which cluster evaporation and nucleation balance, is convected with the particle velocity $v$.

Then, upon eliminating $c_2$ from \eqref{<eq:dtc1>} as discussed above, and using the relations between Fourier modes and hydrodynamic fields (for details see \ref{<B>}), \eqref{<eq:HydrodynamicMomentsS>} -- \eqref{<eq:HydrodynamicMomentsGC>},  equations \eqref{<eq:dts0>} -- \eqref{<eq:dtc2>} give rise to the hydrodynamic equations corresponding to the \emph{canonical model}. In rescaled variables they read:
\begin{subequations}
\begin{eqnarray}
\label{<eq:HydrodynamicEqsTemp1>}
\partial_t\rho_s &=& \rho_c-\left(\rho_s+\rho_c\right)\rho_s,\\
\label{<eq:HydrodynamicEqsTemp2>}
\partial_t\rho_c &=& -\nabla \cdot \vec{g}-\rho_c+\left(\rho_s+\rho_c\right)\rho_s,\\
\label{<eq:HydrodynamicEqsTemp3>}
\partial_t\vec{g} &=& -\nu _1 \vec{g} -\frac{\mu  \kappa }{\nu _2}\vec{g}^2 \vec{g}-\frac{1}{2}\nabla \rho _c+\frac{1}{4\nu _2}\nabla ^2\vec{g}\\
\nonumber
&&+\frac{\zeta _+}{\nu _2}(\vec{g}\cdot \nabla )\vec{g}+\frac{\zeta _-}{\nu _2}\left[(\nabla \cdot \vec{g})\vec{g}-\frac{1}{2}\nabla \left(\vec{g}^2\right)\right]\\
\nonumber
&&+\frac{\mu }{\nu _2^2}\left[\vec{g}(\vec{g}\cdot \grad [\rho_c,\rho_s])-\frac{1}{2}\vec{g}^2\grad[\rho_c,\rho_s]\right]\\
\nonumber
&&+\frac{1}{4\nu _2^2}\left[(\nabla \cdot \vec{g})\grad[\rho_c,\rho_s] - \left(\nabla \vec{g}+\nabla \vec{g}^t\right)\grad[\rho_c,\rho_s]\right],
\end{eqnarray}
\end{subequations}
where
\begin{equation}
\label{<eq:GensityGardient>}
\grad[f,g]=\left(\partial _{f}\nu _2\right)\nabla f+\left(\partial _{g}\nu _2\right)\nabla g,
\end{equation}
and where we have introduced the following abbreviations:
\begin{subequations}
\begin{eqnarray}
\label{<eq:kinetic_coeff1>}
\nu_1 &=& 1-\left(\rho _s-\rho _c\right)+\left(I_{1,0}-2e^{-\sigma ^2/2}I_{1,1}\right)\rho _c,\\
\label{<eq:kinetic_coeff2>}
\nu_2 &=& 1-\left(\rho _s-\rho _c\right)+\left(I_{2,0}-2e^{-2\sigma ^2}I_{1,0}\right)\rho _c,\\
\label{<eq:kinetic_coeff3>}
\mu &=& e^{-2\sigma ^2}-I_{1,0},\\
\label{<eq:kinetic_coeff4>}
\kappa &=& I_{1,0}+I_{2,0}-2e^{-\sigma ^2/2}I_{2,1},\\
\label{<eq:kinetic_coeff5>}
\zeta _{\pm }&=&-\mu \pm \frac{\kappa}{2}.
\end{eqnarray}
\end{subequations}

Equations \eqref{<eq:HydrodynamicEqsTemp1>} -- \eqref{<eq:HydrodynamicEqsTemp3>}  capture the evolution of our \emph{canonical} model system on a hydrodynamic level. More specifically,  \eqref{<eq:HydrodynamicEqsTemp1>} and \eqref{<eq:HydrodynamicEqsTemp2>} describe the spatio-temporal evolution of the particle densities $\rho_s$ and $\rho_c$. Since, by the assumptions underlying our model, no macroscopic flow of \SPS~can build up, only the density of \CPS~($\rho_c$) is subject to convection.
This implies that the genuine hydrodynamic momentum field $\vec g=\vec g_c + \vec g_s \equiv \vec g_c$, is carried solely by the subset of cluster particles.
Therefore we omit the subscript $c$ in \eqref{<eq:HydrodynamicEqsTemp2>} and\eqref{<eq:HydrodynamicEqsTemp3>}
 and denote $\vec g\equiv\vec g_c$.
 The dynamics of both densities is, moreover, driven by source terms, as determined by the reactions discussed in section \ref{sec:ReactionEquations}. The gain and loss parts in these source terms of $\rho_c$ and $\rho_s$ are exactly balanced, such that the total density $\rho=\rho_c+\rho_s$ is conserved.
As an aside we note that any distinction between \SPS~and \CPS~is a purely conceptual matter. Experimentally,  only the total density $\rho$ and the momentum field $\vec g$ are accessible.

\eqref{<eq:HydrodynamicEqsTemp3>}, governing the evolution of the current density $\vec g$, can be interpreted as a generalization of the Navier-Stokes equation to active systems. 
The terms on the right hand side of  \eqref{<eq:HydrodynamicEqsTemp3>} can be given the following interpretation: In the first line, the first two terms account for the local dynamics of $\vec g$. They play a crucial role in establishing and maintaining a state of macroscopic flow, as will be detailed below. The Navier-Stokes equation itself, which conserves momentum, is devoid of these terms. 
In formal analogy to the Navier-Stokes equation, the density gradient in the first line together with the last term in the second line can be interpreted as a pressure gradient. This effective pressure is given by $\frac{1}{2}\left(\rho_c+\frac{\zeta_-}{\nu_2}\,\vec g^2\right)$, when neglecting the density-dependence of $\nu_2$. The last term in the first line is analogous to the shear stress term in the Navier-Stokes equation, with a kinematic viscosity $\sim\nu _2^{-1}$. The second line in  \eqref{<eq:HydrodynamicEqsTemp3>} is a generalization of the convection term to systems not obeying Galilean invariance, where all combinations of $\nabla$ and factors second order in $\vec g$ transforming as vectors are allowed \cite{RamaswamyReview}. Finally, the last two lines describe couplings of the current density $\vec g$ and gradients thereof to density gradients. Note that the density gradients in these coupling terms are all of the same generic structure \eqref{<eq:GensityGardient>}.

\begin{table*}
\centering
		\begin{tabular}{c|c c c c c}
		\hline
\hline

Integral		&		$I_{0,0}$
				&		$I_{1,0}/I_{0,0}$		
				&		$I_{1,1}/I_{0,0}$
				&		$I_{2,0}/I_{0,0}$	
				&		$I_{2,1}/I_{0,0}$		\\
\hline
Value		&		$\frac{8dv(2+\xi)}{3\pi}$
				&		$-\frac{4+\xi}{5(2+\xi)}$
				&		$\frac{3}{16} \frac{8+\pi(\xi-1)}{2+\xi}$
				&		$\frac{6-13\xi}{35(2+\xi)}$
				&		$\frac{3}{16} \frac{\pi(1-\xi)-8}{2+\xi}$\\
\hline
\hline
		\end{tabular}
	\caption{Summary of relevant collision integrals $I_{n,k}$ as a function of the aspect ratio $\xi=L/d$, where  $L$ and $d$ denote particle length and diameter, and where $v$ is the particle velocity. The quantities $I_{n,k}/I_{0,0}$ depend only weakly on the aspect ratio $\xi$. In particular, the signs of $I_{n,k}/I_{0,0}$ do not change with $\xi$, leaving all our present conclusions made on the basis of the kinetic coefficients qualitatively unchanged.}
	\label{<tab:values_of_Ink>}
\end{table*}

As already noted, the \emph{canonical} model equations \eqref{<eq:HydrodynamicEqsTemp1>} -- \eqref{<eq:HydrodynamicEqsTemp3>} conserve the total number of particles. To make this explicit, we define
\begin{subequations}
\label{<eq:RedefRhos>}
\begin{eqnarray}
\label{<eq:RedefRhos1>}
\rho &\equiv& \rho_c + \rho_s,\\
\label{<eq:RedefRhos2>}
\eta &\equiv& \rho_c - \rho_s,
\end{eqnarray}
\end{subequations}
where $\rho$ denotes the overall particle density, and  $\eta$ measures density difference between the two particle classes. The \emph{canonical} model equations then attain the following form:
\begin{subequations}
\begin{eqnarray}
\label{<eq:HydrodynamicEqs1>}
\partial_t\rho &=& -\nabla \cdot \vec{g},\\
\label{<eq:HydrodynamicEqs2>}
\partial_t\eta &=& -\nabla \cdot \vec{g}+\rho^2-(\rho+1)\eta-\rho,\\
\label{<eq:HydrodynamicEqs3>}
\partial_t\vec{g} &=& -\nu_1 \vec{g}-\frac{\mu  \kappa }{\nu _2}\vec{g}^2\vec{g}-\frac{1}{4}\nabla (\rho+\eta)+\frac{1}{4\nu _2}\nabla ^2\vec{g}\qquad\\
\nonumber
&&+\frac{\zeta _+}{\nu _2}(\vec{g}\cdot \nabla )\vec{g}+\frac{\zeta _-}{\nu _2}\left[(\nabla \cdot \vec{g})\vec{g}-\frac{1}{2}\nabla \left(\vec{g}^2\right)\right]\\
\nonumber
&&+\frac{\mu }{\nu _2^2}\left[\vec{g}(\vec{g}\cdot \grad [\rho,\eta])-\frac{1}{2}\vec{g}^2\grad[\rho,\eta]\right]\\
\nonumber
&&+\frac{1}{4\nu _2^2}\left[(\nabla \cdot \vec{g})\grad[\rho,\eta] -   \left(\nabla \vec{g}+\nabla \vec{g}^t\right)\grad[\rho,\eta]\right].
\end{eqnarray}
\end{subequations}
The equation governing $\rho$ expresses the overal conservation of particle number, whereas the source terms of equations \eqref{<eq:HydrodynamicEqsTemp1>} and \eqref{<eq:HydrodynamicEqsTemp2>} combine to determine the local dynamics of the relative density $\eta$ in  \eqref{<eq:HydrodynamicEqs2>}.

Now we turn to the \emph{grand canonical} model, where the \SP~phase is coupled to a particle reservoir, resulting in a situation where \SPS~constitute an isotropic sea of particles which is maintained at a constant density $\rho_{s}^{0}$. Particle number conservation is now violated, and the only non-trivial density dynamics takes place within the phase of \CPS. The hydrodynamic equations corresponding to the \emph{grand canonical} model can be obtained immediately by setting in \eqref{<eq:HydrodynamicEqsTemp1>} -- \eqref{<eq:HydrodynamicEqsTemp3>}  the density of \SPS~to a constant value, yielding:
\begin{subequations}
\begin{eqnarray}
\label{<eq:HydrodynamicEqsTemp1G>}
\rho_s &=& \rho_{s}^{0} = \mbox{const.}\\
\label{<eq:HydrodynamicEqsTemp2G>}
\partial_t\rho_c &=& -\nabla \cdot \vec{g}-\rho_c+\left(\rho_{s}^{0}+\rho_c\right)\rho_{s}^{0},\\
\label{<eq:HydrodynamicEqsTemp3G>}
\partial_t\vec{g} &=& -\nu _1 \vec{g}-\frac{\mu  \kappa }{\nu _2}\vec{g}^2 \vec{g}-\frac{1}{2}\nabla \rho _c+\frac{1}{4\nu _2}\nabla ^2\vec{g}\\
\nonumber
&&+\frac{\zeta _+}{\nu _2}(\vec{g}\cdot \nabla )\vec{g}+\frac{\zeta _-}{\nu _2}\left[(\nabla \cdot \vec{g})\vec{g}-\frac{1}{2}\nabla \left(\vec{g}^2\right)\right]\\
\nonumber
&&+\frac{\mu \partial_{\rho_c} \nu_2}{\nu _2^2}\left[\vec{g}(\vec{g}\cdot \nabla \rho_c)-\frac{1}{2}\vec{g}^2\nabla \rho_c \right]\\
\nonumber
&&+\frac{\partial_{\rho_c} \nu_2}{4\nu _2^2}\left[ (\nabla \cdot \vec{g})\nabla \rho_c - \left(\nabla \vec{g}+\nabla \vec{g}^t\right)\nabla \rho_c  \right].
\end{eqnarray}
\end{subequations}
One final remark is in order: The rescaling scheme introduced in equations \eqref{<equ:dens_scale>} -- \eqref{<equ:dens_scale5>} renders both, the \emph{canonical} and \emph{grand canonical} model equations virtually independent of particle shape. While these equations exhibit a weak dependence on the particles' aspect ratio $L/d$ (via the rescaled collision integrals $I_{n,k}$), this dependence introduces only minor quantitative effects, which are negligible for all present purposes.
To a good approximation we can thus set $L/d=1$ while working with dimensionless variables, and assess the effects entailed by particle shape by restoring original units. Within our present approach, the effects of particle shape are purely quantitative, causing a numerical shift in the characteristic scales, but leaving the qualitative features of the problem 
unaffected. Deep within the ordered phase, i.e. for large densities, we indeed find a qualitative change of the ensuing hydrodynamic instability, as detailed in \ref{<sect:inhomogeneous>}. Nevertheless, this statement has to be taken with a grain of salt because corresponding threshold densities are far beyond the validity of the hydrodynamic equations.



\section{Spatially homogeneous systems}\label{<sect:homogeneous>} 

To investigate the implications of the hydrodynamic equations, 
we start with the simplest case by analyzing spatially homogeneous solutions. These considerations will provide the basis for the study of spatially inhomogeneous systems, which will be the subject of section \ref{<sect:inhomogeneous>}.
Dropping all gradients, the hydrodynamic equations for spatially homogeneous systems for the \emph{canonical} model read
\begin{subequations}
\label{<eq:HydrodynamicEqsHomogeneous>}
\begin{eqnarray}
\label{<eq:HydrodynamicEqsHomogeneous1>}
\partial_t\rho &=& 0  , \\
\label{<eq:HydrodynamicEqsHomogeneous2>}
\partial_t\eta &=&\rho^2-(\rho+1)\eta-\rho, \\
\label{<eq:HydrodynamicEqsHomogeneous3>}
\partial_t\vec{g} &=& -\nu _1\vec g -\frac{\mu  \kappa }{\nu _2}\vec{g}^2\vec{g}.
\end{eqnarray}
\end{subequations}
For the \emph{grand canonical} model we get
\begin{subequations}
\label{<eq:HydrodynamicEqsHomogeneousG>}
\begin{eqnarray}
\label{<eq:HydrodynamicEqsHomogeneous2G>}
\partial_t\rho_c &=&-\rho_c+\left(\rho_{s}^{0}+\rho_c\right)\rho_{s}^{0},\\
\label{<eq:HydrodynamicEqsHomogeneous3G>}
\partial_t\vec{g} &=& -\nu _1\vec g -\frac{\mu  \kappa }{\nu _2}\vec{g}^2\vec{g}.
\end{eqnarray}
\end{subequations}
In both cases, the density dynamics decouples from the momentum current dynamics and can be addressed separately.

In this section, our focus is on the stationary properties of the \emph{canonical} and \emph{grand canonical} model, respectively. While the dynamical approach to the stationary state is model dependent, the system's composition in terms of \SPS~and \CPS, for given total density $\rho$, in the limit $t\rightarrow\infty$ is identical in both cases [refer to  \eqref{<eq:HydrodynamicEqsTemp1>} -- \eqref{<eq:HydrodynamicEqsTemp2>} and \eqref{<eq:HydrodynamicEqsTemp1G>} -- \eqref{<eq:HydrodynamicEqsTemp2G>}]. Since, moreover, the momentum current densities $\vec g$ obey identical dynamical equations, the ensuing analysis of the stationary state is equal for both models.

\subsection{Crossover to clustering}\label{sect:homo_dens}
To assess the density difference between the \CP~and the \SP~phase $\eta$, we calculate the dynamical fixed point $\eta^*$ of \eqref{<eq:HydrodynamicEqsHomogeneous2>}, attracting the dynamics of $\eta(t)$ in the long time limit $t\rightarrow\infty$:
\begin{equation}\label{<eq:fixed_point>}
	\eta^*(\rho) = \frac{\rho^2-\rho}{\rho+1}.  
\end{equation}
\begin{figure}
\centering
\resizebox{0.5\textwidth}{!}{%
  \includegraphics{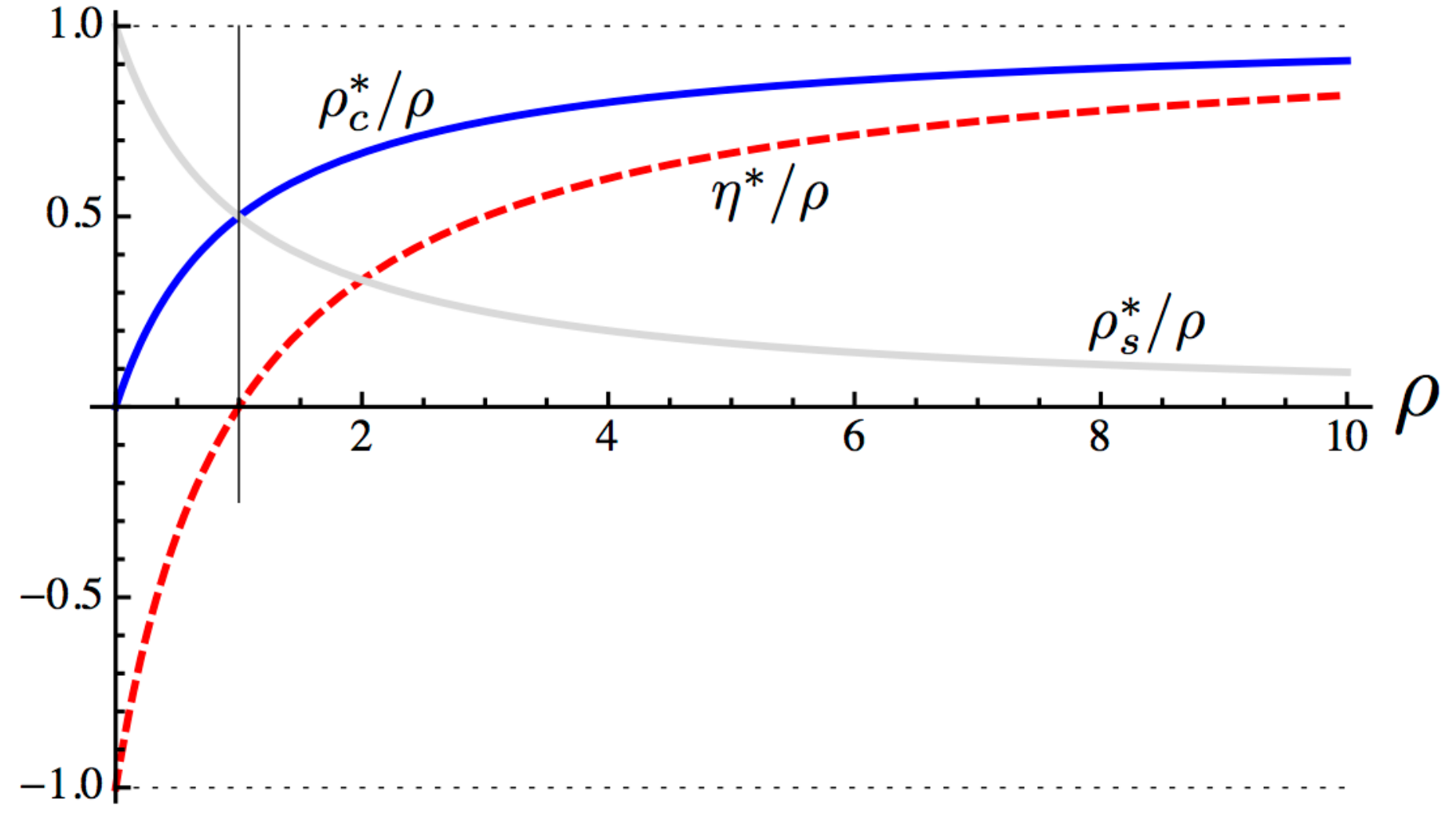}
}
\caption{Fixed points of the homogeneous equations for the \emph{grand canonical} and \emph{canonical} model: The stationary relative density $\eta^*/\rho$, as well as the stationary cluster and single particle density, $\rho_c^*/\rho$ and $\rho_s^*/\rho$, respectively.  The larger $\rho$, the more cluster particles exist in the system. The vertical line corresponds to the density 
$\bar \rho$ above which the number of \CPS~exceeds the number of \SPS.
Note that $\rho_s<1$ holds for all finite values of the total particle density $\rho$. (This is of particular relevance in the context of the \emph{grand canonical} model, where $\rho_s$ is can be considered as control parameter.)}
\label{<fig:Ratio_rho_c_rho_s_delta_c>}
\end{figure} 
The defining equations \eqref{<eq:RedefRhos1>} -- \eqref{<eq:RedefRhos2>} can be used to determine the corresponding (stationary) fixed point densities of \SPS~($\rho_s^*$) and \CPS~($\rho_c^*$) as a function of the total density $\rho$. figure \ref{<fig:Ratio_rho_c_rho_s_delta_c>} summarizes these findings: Upon increasing the total density $\rho$, the ratio $\eta^*/\rho$ continuously grows from $\eta^*/\rho=-1$ at $\rho=0$, asymptotically approaching $\eta^*/\rho=1$ as $\rho\rightarrow\infty$. Based on the sign of $\eta^*$, two density regimes can be distinguished: In the low density regime ($\rho\ll1$, $\eta^*<0$) particle collisions, underlying the formation of clusters, occur at much smaller rates than cluster evaporation events. Only a small fraction of all particles organize themselves in clusters leading to a relatively dense population of \SPS~and correspondingly small density of \CPS. In the high density regime ($\rho\gg1$, $\eta^*>0$), the situation is reversed: Large overall densities imply frequent particle collisions and, consequently,  cluster formation and cluster growth dominate over cluster evaporation. In this regime, the number of \CPS~exceeds the number of \SPS.

The crossover between the \SP~dominated low density regime and the \CP~dominated high density regime occurs at the crossover density $\bar\rho=\hat\rho_b=1$, where both, the \SP~and the \CP~populations are of equal size [i.e. $\eta^*(\bar\rho)=0$]. The relation between the crossover density to clustering, and the geometrical shape of the constituent particles has been addressed previously in Ref. \cite{Peruani_rods}, based on agent-based simulations and a mean-field type analytical analysis. Using our definition of the crossover density $\bar\rho$ we can establish the corresponding relation simply by restoring original units [equation \eqref{<rho_b>}]. Using packing fraction $\bar p\simeq\bar\rho\,L\,d$ instead of particle density, and assuming for the sake of simplicity $L/d\gg1$, which allows us to estimate the particle surface $A_0\simeq Ld$, we find:
\begin{equation}
\label{<eq:PvsXi>}
\bar p \simeq \frac{\epsilon\,Ld}{I_{0,0}} = \frac{3\pi}{8v}\frac{\epsilon L}{2+L/d},
\end{equation}
which correctly reproduces the findings of Ref. \cite{Peruani_rods} (taking into account that the cluster evaporation rate is assumed to be proportional to the inverse particle length, $\epsilon\propto L^{-1}$). 
For the sake of completeness, we note that the definition of the clustering crossover density in reference \cite{Peruani_rods} is based on the cluster size distribution, and thus does not necessarily coincide with our definition. We stress, however, that in our description the scaling structure in  equation \eqref{<eq:PvsXi>} is completely generic. It is an immediate consequence of the characteristic scales of our model and of the fact that the rescaled hydrodynamic model equations are (virtually) independent of particle shape. The structure of equation \eqref{<eq:PvsXi>} is thus robust under an arbitrary redefinition of the (rescaled) crossover density $\bar\rho$.

\subsection{Homogeneous equations for momentum current density}\label{<sect:homo_pol>}

Having examined the composition of the system in terms of \SP~and \CP~densities, we now turn to a discussion of the spatially homogeneous solutions for the momentum  current density $\vec g$. Due to rotational invariance of  \eqref{<eq:HydrodynamicEqsHomogeneous3>}, only the magnitude $g=|\vec g|$ of the momentum current density, but not its direction, evolves in time.  We can thus concentrate on the scalar equation
\begin{equation}
\label{<eq:g_scalar>}
\partial_t g = -\nu _1 g -\frac{\mu  \kappa }{\nu _2}\,g^3,
\end{equation}
which leads to the following fixed points $g^*$ as the attractor of the dynamics of $g$ in the limit of long times:
\begin{equation}
\label{<eq:g_scalar_FPs>}
g^*=
\begin{cases}
0 &for  \nu_1>0,\\
g_0=\sqrt{-\frac{\nu_1\nu_2}{\mu\kappa}}&for \nu_1<0.
\end{cases}
\end{equation}
It can be shown, that the coefficient in front of the cubic term in \eqref{<eq:g_scalar>} is indeed strictly positive  for all control parameters of density $\rho$ and noise $\sigma$ consistent with $\nu_1<0$, ensuring the existence of the non-trivial fixed point in the second line of \eqref{<eq:g_scalar_FPs>}.

Depending on the sign of the linear coefficient $\nu_1$, two parameter regimes can thus be distinguished: Parameters leading to $\nu_1>0$ render stable an overall homogeneous and isotropic state with vanishing macroscopic flow $\vec g = 0$. Upon crossing the \emph{phase boundary}
\begin{equation}
\label{<eq:hypersfc>}
\nu_1(\rho,\sigma) = 0
\end{equation}
in parameter space, the isotropic solution gets unstable and a macroscopic current density of non-zero amplitude builds up. In equation \eqref{<eq:hypersfc>} we used that the density difference $\eta$, in the stationary limit, is a function of the total density $\rho$; cf. equation \eqref{<eq:fixed_point>}. Hence, in the limit of long times, $\nu_1$ is a function of the total density $\rho$ and the noise parameter $\sigma$, only.

\begin{figure}
\centering
\resizebox{0.5\textwidth}{!}{%
  \includegraphics{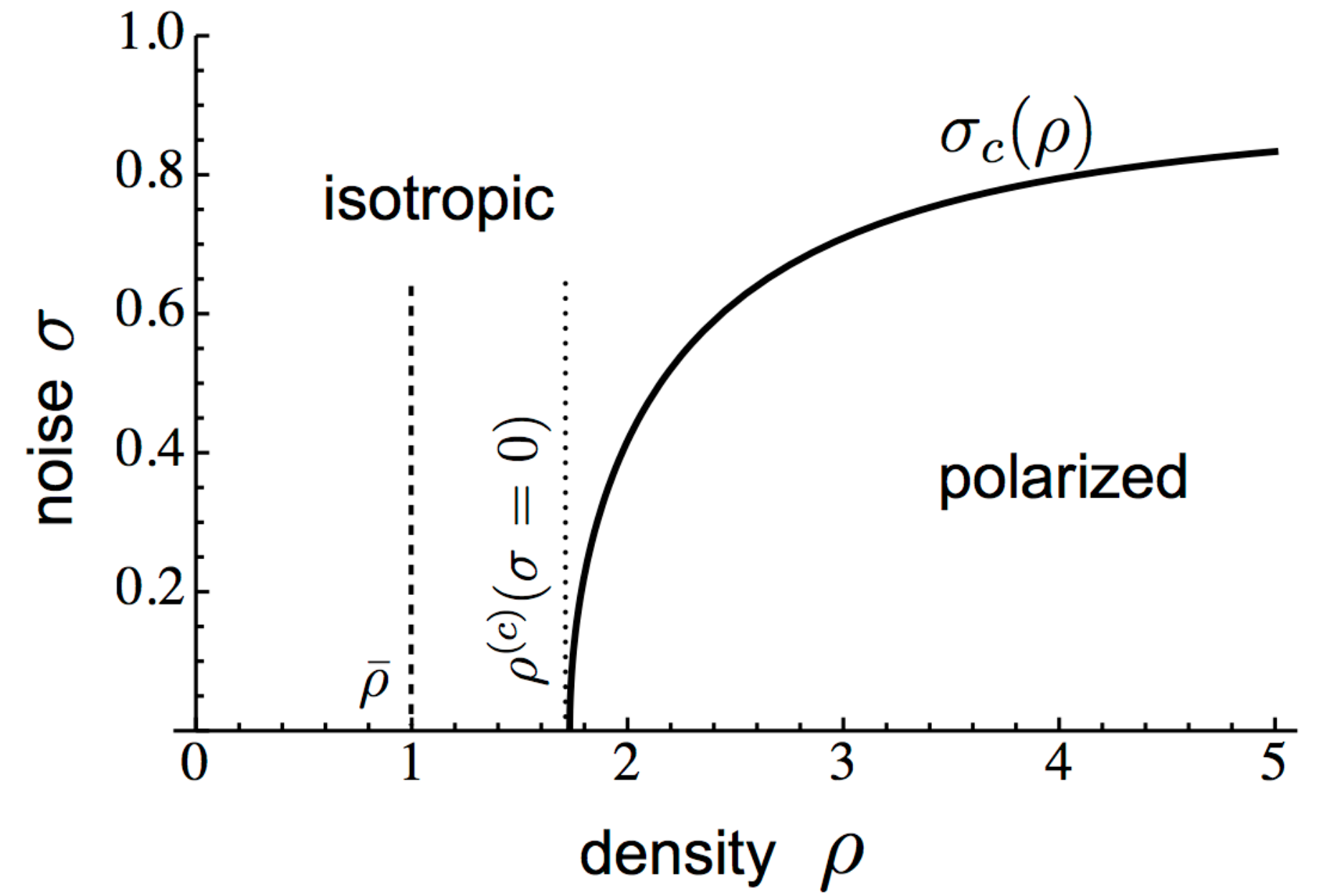}
}
\caption{Phase diagram given by the homogeneous equations for the \emph{canonical}  and \emph{grand canonical} model. For noise values smaller than the critical value, $\sigma<\sigma_c$, the isotropic state becomes unstable, giving rise to a state of collective motion of non-zero macroscopic momentum current. For $\sigma>\sigma_c$ the isotropic state ($g_0=0$) represents a stable solution. The vertical dotted line indicates the transition density $\rho^{(c)}$ at zero collision noise $\sigma = 0$, and the vertical dashed line corresponds to the crossover density $\bar \rho$, above which the number of \CPS~exceeds the number of \SPS.}
\label{<fig:phasediagramm_homogen>}      
\end{figure}

Using the definition of the coefficient $\nu_1$, equation \eqref{<eq:kinetic_coeff1>}, we can readily calculate the shape of the phase boundary in the $\sigma$$\rho$ --plane:
\begin{equation}\label{<equ:sigma_C>}
	\sigma_c(\rho) = \sqrt{-2 \, \ln{\left(\frac{2}{3} + \rho^{-2}\right)}},\qquad(\rho \geq \sqrt{3}),
\end{equation}
where we used $I_{1,0}=-\frac{1}{3}$ and $I_{1,1}=\frac{1}{2}$.
The corresponding phase diagram is shown in figure \ref{<fig:phasediagramm_homogen>}.

To conclude this section, we note that the analysis of spatially homogeneous systems corroborates the general physical picture of active systems, that was alluded to in the introduction (e.g. cf. reference \cite{Gregoire:2004uf}): Even in the absence of noise, $\sigma = 0$, for which the threshold density $\rho^{(c)}$ is lowest, the fully isotropic state $\vec g = 0$ remains stable up to a critical density $\rho^{(c)}(\sigma=0)=\sqrt{3}\,\bar\rho$, which lies well beyond the density $\bar\rho$ indicating the crossover to clustering. We thus extract the following physical picture; cf.  figure \ref{<fig:phasediagramm_homogen>}: For low densities, $\rho<\bar\rho$, cluster evaporation dominates over cluster assembly via particle collisions and clusters form only transiently. The system most closely resembles a structureless, isotropic ``sea of particles''. At intermediate densities, $\bar\rho<\rho<\rho^{(c)}$, particle collisions are more frequent. The emergence of clusters is now a virtually  persistent phenomenon, with cluster evaporation occurring at a lower rate than cluster formation and growth. Yet, the collision rates between clusters (i.e. collisions among \CPS) are still too low to orchestrate macroscopic order, leading to an overall isotropic ``sea of clusters''. Finally, for large densities, $\rho>\rho^{(c)}$, the frequency of collisions among clusters is high enough to establish collective motion even on macroscopic scales.

\section{Stability of inhomogeneous hydrodynamic equations}\label{<sect:inhomogeneous>} 

From our hitherto discussions, we have ascertained that the isotropic, homogeneous state ($\rho=\mbox{const.}$ and $\vec g = 0$) becomes unstable for sufficiently large densities. 
 Yet, from a purely homogeneous analysis we cannot tell anything about the spatial structure of such a macroscopic broken-symmetry state. Nor can we be sure that the isotropic and homogeneous solution for $\rho<\rho^{c}(\sigma)$ is indeed stable with respect to spatially inhomogeneous perturbations. In this section, we therefore test the linear stability of the homogeneous isotropic and non-isotropic base states with respect to wavelike perturbations of arbitrary wave number. Unlike the homogeneous model equations, the full hydrodynamic model equations are different for both, the \emph{canonical} and \emph{grand canonical} model, implying different dispersion relations describing the growth of such wave-like perturbations. We will thus analyze both models separately, and show that particle conservation does indeed influence pattern formation in essential respects.

\subsection{Linearization about stationary, spatially homogeneous base states}
We start by linearizing the hydrodynamic equations for the \emph{canonical} model. In the \emph{canonical} model, the total number of particles is conserved, and the appropriate base state reads (cf. section \ref{<sect:homogeneous>})
\begin{subequations}
\begin{eqnarray}
\label{<eq:BaseCanonical1>}
\rho &=& \rho_h = \mbox{const.},\\
\label{<eq:BaseCanonical2>}
\eta &=& \eta^*(\rho_h) = \frac{\rho_h^2-\rho_h}{\rho_h+1},\\
\label{<eq:BaseCanonical3>}
\vec g &=& \vec{g}_h \in \left\{0,\;g_0=\sqrt{-\frac{\nu_1\nu_2}{\mu\kappa}}\right\}\hat{\vec e}_g,
\end{eqnarray}
\end{subequations}
where $\hat{\vec e}_g$ denotes the unit vector in the direction of the homogeneous polarization, and where all fields of the base states are assumed to be constant both in space and time.
We are going to investigate the linear stability of the solutions \eqref{<eq:BaseCanonical1>} -- \eqref{<eq:BaseCanonical3>} against wave-like perturbations, employing the following ansatz:
\begin{subequations}
\begin{eqnarray}
\label{<eq:PertAnsatzC1>}
\rho(\vec x,t) &=& \rho_{h} + \delta \rho(\vec x, t),\\
\label{<eq:PertAnsatzC2>}
\eta(\vec x,t) &=& \eta^{*} + \delta \eta(\vec x, t),\\
\label{<eq:PertAnsatzC3>}
\vec{g}(\vec x,t) &=& \vec{g}_{h} + \delta \vec{g}(\vec x, t),
\end{eqnarray}
\end{subequations}
where the perturbations are plane waves
\begin{subequations}
\begin{eqnarray}
\label{<eq:PertAnsatzC4>}
\delta \rho (\vec{x},t) &=& \delta \rho_0\,e^{st+i \vec{q}\cdot \vec{x}},\\
\label{<eq:PertAnsatzC5>}
\delta \eta (\vec{x},t) &=& \delta \eta_0 \,e^{st+i \vec{q}\cdot \vec{x}},\\
\label{<eq:PertAnsatzC6>}
\delta \vec{g} (\vec{x},t) &=& \delta \vec{g}_0\,e^{st+i \vec{q}\cdot \vec{x}}.
\end{eqnarray}
\end{subequations}
In the equations above, $\vec q$ denotes the wave vector and $s$ is the growth rate.
Inserting this ansatz into the hydrodynamic equations \eqref{<eq:HydrodynamicEqs1>} -- \eqref{<eq:HydrodynamicEqs3>}, we obtain the following eigenvalue problem:
\begin{subequations}
\begin{eqnarray}
\label{<equ:canonical_linear1>}
s\,\delta \rho_0 &=& -i \vec q\cdot \delta\vec{g}_0,\\
\label{<equ:canonical_linear2>}
s\,\delta\eta_0 &=& \left(2\rho_h-\eta^*-1\right)\delta \rho_0 -\left(1+\rho _h\right)\delta\eta_0-i \vec q\cdot \delta\vec{g}_0,\\
\label{<equ:canonical_linear3>}
s\,\delta\vec{g}_0 &=& \left[\frac{\partial_{\rho}\nu_2\,\mu }{\nu _2^2}\left(\kappa\,\vec{g}_h^2\,\vec{g}_h+\left(\vec{g}_h\cdot i\vec{q}\right)\vec{g}_h-\frac{i\vec{q}}{2}\,\vec{g}_h^2\right)-\partial_{\rho}\nu_1\,\vec{g}_h-\frac{i\vec{q}}{4}\right]\delta \rho\\
\nonumber
&& +\left[\frac{\partial_{\eta}\nu_2\,\mu }{\nu _2^2}\left(\kappa\,\vec{g}_h^2\,\vec{g}_h+\left(\vec{g}_h\cdot i\vec{q}\right)\vec{g}_h-\frac{i\vec{q}}{2}\,\vec{g}_h^2\right)-\partial_{\eta}\nu_1\,\vec{g}_h-\frac{i\vec{q}}{4}\right]\delta \eta\\
\nonumber
&& +\left[\frac{\zeta _+}{\nu _2}\left(i\vec{q}\cdot \vec{g}_h\right)-\frac{\vec{q}^2}{4\nu _2}-\nu_1-\frac{\mu\kappa }{\nu _2}\vec{g}_h^2\right]\delta\vec{g}_0-\frac{2\mu  \kappa }{\nu _2}\vec{g}_h\left(\vec{g}_h\cdot \delta\vec{g}_0\right)\\
\nonumber
&&+\frac{\zeta _-}{\nu _2}\Bigl[\vec{g}_h(i\vec{q}\cdot \delta\vec{g}_0\vec{)}-i\vec{q}\left(\vec{g}_h\cdot \delta\vec{g}_0\right)\Bigr].
\end{eqnarray}
\end{subequations}

Unlike the \emph{canonical} model, the \emph{grand canonical} model conserves the number of \SPS, but not the total number of particles. The appropriate base state in this case reads: 
\begin{subequations}
\begin{eqnarray}
\label{<eq:BaseGCanonical1>}
\rho_s &=& \mbox{const.},\\
\label{<eq:BaseGCanonical2>}
\rho_c &=& \rho_c^*(\rho_s) = \frac{\rho_s^2}{1-\rho_s},\\
\label{<eq:BaseGCanonical3>}
\vec g &=& \vec{g}_h \in \left\{0,\;g_0=\sqrt{-\frac{\nu_1\nu_2}{\mu\kappa}}\right\}\hat{\vec e}_g.
\end{eqnarray}
\end{subequations}
We investigate the linear stability of these solutions, using a perturbation ansatz analogous to equations. \eqref{<eq:PertAnsatzC1>} -- \eqref{<eq:PertAnsatzC6>}:
\begin{subequations}
\begin{eqnarray}
\label{<eq:PertAnsatzGC1>}
\rho_c(\vec x,t) &=& \rho_{c}^* + \delta \rho_c(\vec x, t),\\
\label{<eq:PertAnsatzGC2>}
\vec{g}(\vec x,t) &=& \vec{g}_{h} + \delta \vec{g}(\vec x, t),
\end{eqnarray}
\end{subequations}
with
\begin{subequations}
\begin{eqnarray}
\label{<eq:PertAnsatzGC3>}
\delta \rho_c(\vec{x},t) &=& \delta \rho_{c}^{0}\,e^{st+i \vec{q}\cdot \vec{x}},\\
\label{<eq:PertAnsatzGC4>}
\delta \vec{g} (\vec{x},t) &=& \delta \vec{g}_0\,e^{st+i \vec{q}\cdot \vec{x}}.
\end{eqnarray}
\end{subequations}
Inserting this ansatz into equations \eqref{<eq:HydrodynamicEqsTemp1G>} -- \eqref{<eq:HydrodynamicEqsTemp3G>}, we obtain:
\begin{subequations}
\begin{eqnarray}
\label{<eq:LinearizedEqsGC1>}
s\,\delta\rho _c^0&=&\left(\rho _s-1\right)\delta\rho_c^0-i\vec{q}\cdot \delta\vec{g}_0\\
\label{<eq:LinearizedEqsGC2>}
s\,\delta\vec{g}_0&=&\Bigg[\frac{\partial_{\rho_c}\nu_2\,\mu }{\nu _2^2}\left(\kappa\,\vec{g}_h^2\,\vec{g}_h+\left(\vec{g}_h\cdot i\vec{q}\right)\vec{g}_h-\frac{i\vec{q}}{2}\vec{g}_h^2\vec{ }\right)\\
\nonumber
&&-\partial_{\rho_c}\nu_1\vec{g}_h-\frac{i\vec{q}}{2}\Bigg]\delta\rho_c^0\\
\nonumber
&&+\left[\frac{\zeta _+}{\nu _2}\left(i\vec{q}\cdot \vec{g}_h\right)-\frac{\vec{q}^2}{4\nu _2}-\nu_1-\frac{\mu\kappa }{\nu _2}\vec{g}_h^2\right]\delta\vec{g}_0-\frac{2\mu  \kappa }{\nu _2}\vec{g}_h\left(\vec{g}_h\cdot \delta\vec{g}_0\right)\\
\nonumber
&&+\frac{\zeta _-}{\nu _2}\Bigl[\vec{g}_h(i\vec{q}\cdot \delta\vec{g}_0\vec{)}-i\vec{q}\left(\vec{g}_h\cdot \delta\vec{g}_0\right)\Bigr].
\end{eqnarray}
\end{subequations}

\subsection{Stability of the disordered state $g_0=0$}\label{<sec:StabDisorderedState>}

\begin{figure}
\centering
\resizebox{1.0\textwidth}{!}{%
  \includegraphics{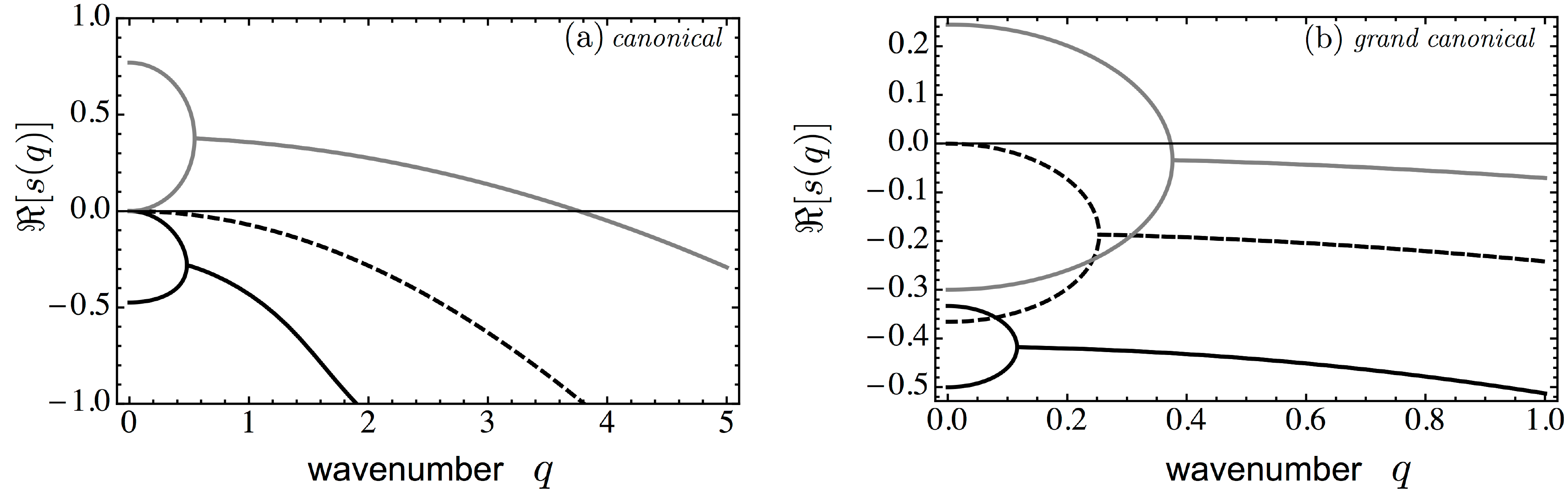}
}
\caption{Fastest growth rate of $\Re[s(q)]$ as a function of of the wave number $q=|\vec q|$ for the \emph{canonical} (a) and \emph{grand canonical} model (b), each for $\sigma=0$. The disordered state is stable for all wave numbers $q$ if $\rho<	\rho^{(c)}(\sigma)$. The marginal case, $\rho = \rho^{(c)}(\sigma)$, is dashed. An instability (grey) occurs for densities larger than the corresponding homogeneous critical density $\rho^{(c)}(\sigma)$. Similar behavior is found for $\sigma\not=0$.}
\label{<fig:Spatial_analysis_homogen>}
\end{figure}

We start by considering the homogeneous and isotropic base state, which was shown to be stable against spatially homogeneous perturbations for $\rho < \rho^{(c)}(\sigma)$; cf. section \ref{<sect:homo_pol>}. To assess the stability of this state with respect to perturbations of arbitrary (non-zero) wavevectors in the \emph{canonical} model, we use the linearized hydrodynamic equations \eqref{<equ:canonical_linear1>} -- \eqref{<equ:canonical_linear3>} with $g_h=0$. The resulting eigenvalue problem is most conveniently expressed in matrix form:
\begin{equation}
\label{<eq:MatrixC0>}
s
\begin{pmatrix}
\delta\rho_0\\
\delta\eta_0\\
\delta g_0
\end{pmatrix}
=
\begin{pmatrix}
0 & 0 & -iq\\
2\rho_h-\eta^*-1 & -(1+\rho_h) & -iq\\
-iq/4 & -iq/4 & -\nu_1-q^2/(4\nu_2)
\end{pmatrix}
\begin{pmatrix}
\delta\rho_0\\
\delta\eta_0\\
\delta g_0
\end{pmatrix}.
\end{equation}
The corresponding eigenvalue problem for the \emph{grand canonical} model is found from equations \eqref{<eq:LinearizedEqsGC1>} -- \eqref{<eq:LinearizedEqsGC2>}, and attains the following form:
\begin{equation}
\label{<eq:MatrixGC0>}
s
\begin{pmatrix}
\delta\rho_c^0\\
\delta g_0
\end{pmatrix}
=
\begin{pmatrix}
(\rho_s-1) & -iq\\
-iq/2 & -\nu_1-q^2/(4\nu_2)
\end{pmatrix}
\begin{pmatrix}
\delta\rho_c^0\\
\delta g_0
\end{pmatrix}.
\end{equation}
For $\vec{g}_h=0$,  \eqref{<equ:canonical_linear3>} or \eqref{<eq:LinearizedEqsGC2>}, respectively, implies $\vec q\,||\,\delta\vec g_0$, allowing to replace the vectors $\vec q$ and $\delta\vec g_0$ by their respective magnitudes $q$ and $\delta g_0$. We solved both eigenvalue problems numerically, for arbitrary wavenumbers $q>0$ with the results shown in figure \ref{<fig:Spatial_analysis_homogen>}. Note that in both models the real parts of all eigenvalues are negative for all wavenumbers $q>0$, provided the particle density $\rho<\rho^{(c)}$. The spatially homogeneous, isotropic state is thus stable against small perturbations with \emph{arbitrary} wavevectors.

For densities $\rho>\rho^{(c)}$, in contrast, a narrow band of  positive eigenvalues emerges in both models, located at wavenumbers $q\ll1$. Equations \eqref{<eq:MatrixC0>} and \eqref{<eq:MatrixGC0>} evaluated at $q=0$ return nothing but the linearized versions of the homogeneous hydrodynamic equations,  \eqref{<eq:HydrodynamicEqsHomogeneous1>} -- \eqref{<eq:HydrodynamicEqsHomogeneous3>} and \eqref{<eq:HydrodynamicEqsHomogeneous2G>} -- \eqref{<eq:HydrodynamicEqsHomogeneous3G>}. To gain new insights, we will therefore examine the limit $q\rightarrow0$ and consider the eigenvalues of the above coefficient matrices to leading order in the wavenumber $q$.

In this limit of small wavenumbers, the \emph{grand canonical} coefficient matrix, given in \eqref{<eq:MatrixGC0>}, approaches diagonal form and the dynamics of density fluctuations $\delta\rho_c^{0}$ and momentum current density fluctuations $\delta g_0$ practically decouple. Since $\rho_s<1$ (cf.  figure \ref{<fig:Ratio_rho_c_rho_s_delta_c>}), the first eigenvalue $s_1^{(\mbox{\tiny GC})}=\rho_s-1+\mathcal{O}(q^2)$ is strictly negative and density fluctuations decay exponentially. The second eigenvalue, $s_2^{(\mbox{\tiny GC})}=-\nu_1+\mathcal{O}(q^2)$, is positive at small wavenumbers leading to an instability in the momentum current density against long wavelength fluctuations.

In the case of the \emph{canonical} model \eqref{<eq:MatrixC0>}, the coefficient matrix approaches block diagonal form in the limit of small wavenumbers. Again, the dynamics of momentum current density fluctuations $\delta g_0$ practically decouples from density fluctuations ($\delta\rho_0$ and $\delta\eta_0$), with momentum current density fluctuations being amplified by virtue a positive eigenvalue $s_3^{(\mbox{\tiny C})}=-\nu_1+\mathcal{O}(q^2)$ at small wavenumbers. In contrast to the \emph{grand canonical} model, however, particle conservation entails a marginally stable mode $s_1^{(\mbox{\tiny C})}(q=0)=0$, which turns positive for $q\gtrsim0$: $s_1^{(\mbox{\tiny C})}\propto q^2$ (the remaining eigenvalue $s_2^{(\mbox{\tiny C})}=-(1+\rho_h)+\mathcal{O}(q^2)$ is strictly negative).

To sum up, the study of the linear stability of the homogeneous, isotropic state against spatially inhomogeneous perturbations of arbitrary wave vectors strongly suggests that particle conservation plays a vital role in the context of pattern formation. Both models exhibit spontaneous symmetry breaking by establishing a state of macroscopic collective motion. In the canonical model, in addition, conservation of total particle number entails a marginally stable density mode at $q=0$ which is absent in the \emph{grand canonical} model. This mode, in turn, gives rise to a density instability at small, non-zero wavenumbers, accompanying the spontaneous symmetry breaking event for $\rho>\rho^{(c)}$. We note, however, that, at this point of the discussions, the existence of a narrow band of unstable modes at small wavenumbers does not allow for any conclusions concerning the structure of the macroscopic density and momentum current density  for $\rho>\rho^{(c)}$. We will address this issue in greater detail in the following section.

\subsection{Stability of the broken symmetry state $g_0 > 0$}
\label{<sec:InhomStabgg0>}
Both, the \emph{canonical} and \emph{grand canonical} model exhibit spontaneous symmetry breaking for overall densities $\rho>\rho^{(c)}(\sigma)$. To illuminate the spatial structure of this broken symmetry state, we start from the most simple case of a  spatially homogeneous state of collective motion, and examine its stability with respect to wavelike perturbations in the hydrodynamic particle and momentum current densities. Without loss of generality, we assume the direction of the macroscopic momentum current density to coincide with the $x$-direction and choose $\vec g_h = g_0\,\hat{\vec e}_x$. The wave vector $\vec q$ of the perturbation fields is assumed to make an angle $\psi$ with the macroscopic momentum current density $\vec g_h$, yielding  $\psi= \angle (\vec q, \vec e_x)$ and $\vec q=q\,(\cos{(\psi)},\sin{(\psi)})$ with $q=|\vec q|$.

The linearized \emph{canonical} model equations \eqref{<equ:canonical_linear1>} -- \eqref{<equ:canonical_linear3>} then attain the following form:
\begin{subequations}
\begin{eqnarray}
\label{<eq:canonical_linear1_Inhomogen>}
s\,\delta \rho_0 &=& -i q \cos(\psi)\delta g_{x,0}-i q \sin(\psi)\delta g_{y,0},\\
s\,\delta\eta_0 &=& \left(2\rho_h-\eta^*(\rho_h)-1\right)\delta \rho_0 -\left(1+\rho _0\right)\delta\eta_0\qquad\\
\nonumber
&&-i q \cos(\psi)\delta g_{x,0}-i q \sin(\psi)\delta g_{y,0},\\
s\,\delta g_{x,0} &=& \Biggl[\frac{1}{2}i q \cos(\psi)\left( g_0^2\,\frac{\mu}{\nu _2^2}\partial _{\rho }\nu _2-\frac{1}{2}\right)-g_0\left(\partial _{\rho }\nu_1-\frac{\mu\kappa }{\nu_2^2}g_0^2\partial _{\rho}\nu_2\right)\Biggr]\delta \rho_0\\
\nonumber
&+&\Biggl[\frac{1}{2}i q \cos(\psi)\left( g_0^2\,\frac{\mu}{\nu _2^2}\partial _{\eta}\nu _2-\frac{1}{2}\right)-g_0\left(\partial_{\eta}\nu_1-\frac{\mu\kappa }{\nu_2^2}g_0^2\partial_{\eta}\nu _2\right)\Biggr]\delta\eta_0\\
\nonumber
&&+\left(i q \cos(\psi)\frac{\zeta _+}{\nu _2}g_0-\frac{q^2}{4\nu _2} -\frac{2\mu\kappa}{\nu_2}g_0^2\right)\delta g_{x,0}\\
\nonumber
&&+i q\frac{\zeta _-}{\nu _2}\sin(\psi)g_0\delta g_{y,0},\\
\label{<eq:canonical_linear4_Inhomogen>}
s\,\delta g_{y,0} &=&-\frac{1}{2}i q \sin(\psi)\left( g_0^2\,\frac{\mu}{\nu _2^2}\partial _{\rho }\nu _2+\frac{1}{2}\right)\delta \rho_0 \quad\\
\nonumber
&&-\frac{1}{2}i q \sin(\psi)\left( g_0^2\,\frac{\mu}{\nu _2^2}\partial _{\eta}\nu _2+\frac{1}{2}\right)\delta\eta_0\\
\nonumber
&&-i q \sin(\psi)\frac{\zeta _-}{\nu _2}\,g_0\,\, \delta g_{x,0}\\
\nonumber
&&+\left(i q \cos(\psi)\frac{\zeta _+}{\nu _2}g_0-\frac{q^2}{4\nu _2}\right)\delta g_{y,0}.
\end{eqnarray}
\end{subequations}
The corresponding equations for the \emph{grand canonical} model read:
\begin{subequations}
\begin{eqnarray}
\label{<eq:dens_dynamics_GC1>}
s\, \delta\rho_{c}^{0} &=&  \left(\rho_{s}-1\right)\delta\rho_{c}^{0} -i q \cos(\psi)\delta g_{x,0}-i q \sin(\psi)\delta g_{y,0},\\
s\, \delta g_{x,0} &=& \Biggl[\frac{1}{2}i q \cos(\psi)\left( g_0^2\,\frac{\mu}{\nu _2^2}\partial _{\rho_c}\nu _2-1\right)\\
\nonumber
&-&g_0\left(\partial_{\rho_c}\nu_1-\frac{\mu\kappa }{\nu_2^2}g_0^2\partial_{\rho_c}\nu _2\right)\Biggr]\delta\rho_{c}^{0}\\
\nonumber
&&+\left(i q \cos(\psi)\frac{\zeta _+}{\nu _2}g_0-\frac{q^2}{4\nu _2} -\frac{2\mu\kappa}{\nu_2}g_0^2\right)\delta g_{x,0}\\
\nonumber
&&+i q\frac{\zeta _-}{\nu _2}\sin(\psi)g_0\, \, \delta g_{y,0},\\
\label{<eq:dens_dynamics_GC3>}
s\, \delta g_{y,0} &=&-\frac{1}{2}i q \sin(\psi)\left( g_0^2\,\frac{\mu}{\nu _2^2}\partial _{\rho_{c}}\nu _2+1\right)\delta\rho_{c}^{0}\\
\nonumber
&&-i q \sin(\psi)\frac{\zeta _-}{\nu _2}\,g_0\,\, \delta g_{x,0}\\
\nonumber
&&+\left(i q \cos(\psi)\frac{\zeta _+}{\nu _2}g_0-\frac{q^2}{4\nu _2}\right)\delta g_{y,0}.
\end{eqnarray}
\end{subequations}
In equations \eqref{<eq:canonical_linear1_Inhomogen>} -- \eqref{<eq:dens_dynamics_GC3>} we used $-\nu_1-\frac{\mu\kappa}{\nu_2}\,g_0^2=0$, which directly follows from the definition of $g_0$, given in  equation \eqref{<eq:g_scalar_FPs>}. We numerically solved both eigenvalue problems in the immediate vicinity of the ordering transition line $\rho=\rho^{(c)}(\sigma)$.

\begin{figure}
\centering
\resizebox{1.0\textwidth}{!}{%
  \includegraphics{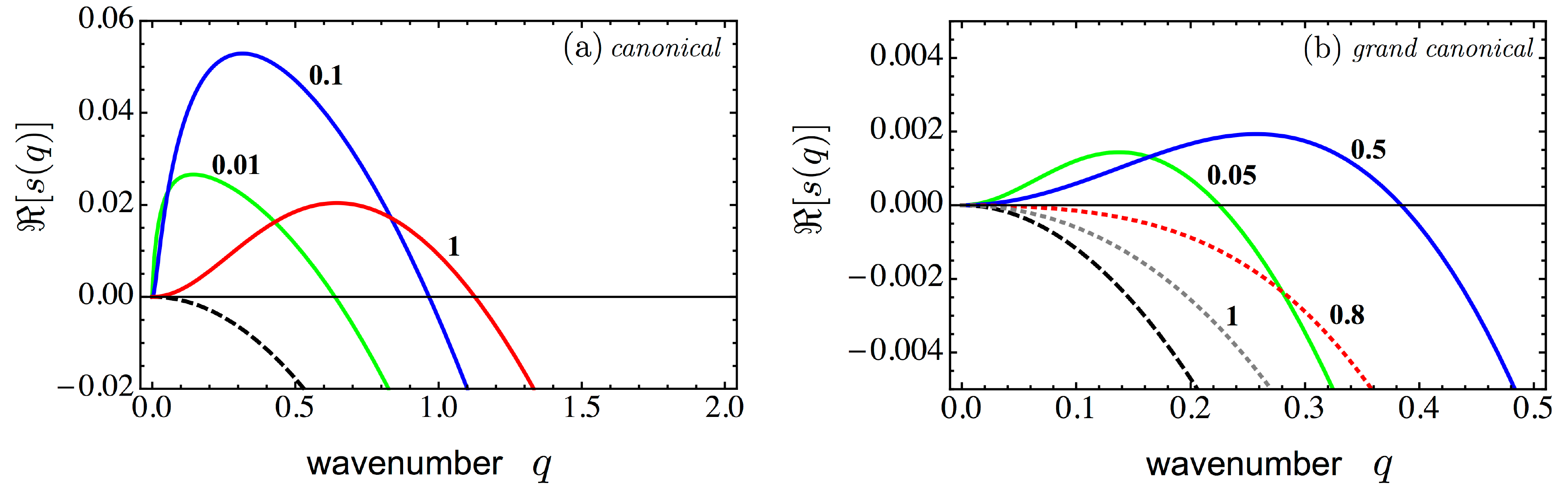}
}
\caption{Largest growth rate of $\Re[s(q)]$ as a function of the wave number $q$ for $\sigma=0$ and several values for the total particle density $\rho$. The marginal $\rho = \rho^{(c)}$ is dashed. Further values are $\rho= \rho^{(c)} +\Delta$ with $\Delta$ indicated in the figure.  
(a) \emph{Canonical model:} 
For $\rho > \rho^{(c)}$, longitudinal perturbations ($\psi=0$) are unstable.
 (b) \emph{Grand canonical model:} Transversal ($|\psi|=\pi/2$) perturbations are unstable closely above the  critical density  $\rho^{(c)}$ (refer to green and blue curve corresponding to $\Delta\in \{0.05,0.5\}$). For larger densities, i.e. 
 $\Delta> 0.7$ for $\sigma=0$, the transversal instability re-stabilizes again (dotted curves, $\Delta\in \{0.8,1\}$). However, the density regime hosting this transversal instability vanishes completely for noise values larger than $\sigma_r$, as illustrated in \ref{<fig:Spatial_homogen_with_trans_GC>}.}
\label{<fig:Spatial_analysis_C_Long_GC_trans>}       
\end{figure}

In the case of the \emph{canonical} model, we find that the most unstable mode occurs for longitudinal perturbations, i.e. perturbations with wave vectors parallel to the direction of macroscopic motion, $\vec q\,||\,\vec g_0$ ($\psi=0$). \ref{<fig:Spatial_analysis_C_Long_GC_trans>}a shows the corresponding eigenvalues as functions of the wavenumber $q$ for a set of density values slightly beyond $\rho=\rho^{(c)}$. Further inspection of the coupling coefficients in equations \eqref{<eq:canonical_linear1_Inhomogen>} -- \eqref{<eq:canonical_linear4_Inhomogen>} reveals that this longitudinal instability only affects the amplitude of $\vec g$ leaving the direction unchanged: For $\psi = 0$, the dynamics of $\delta g_{y,0}$ decouples and momentum current density  fluctuations perpendicular to the direction of macroscopic motion decay exponentially,
\begin{equation}
\delta g_{y,0} = s_4^{(\mbox{\tiny C})}\,\delta g_{y,0},
\end{equation}
with a rate
\begin{equation}
\Re\left[s_4^{(\mbox{\tiny C})}\right] = -\frac{q^2}{4\nu _2} < 0,
\end{equation}
which approaches zero for $q\rightarrow0$, as expected for a broken symmetry variable. To assess the nature of the instability in greater detail, we calculated the  eigenvector corresponding to the most unstable longitudinal mode (evaluated at the most unstable wavenumber). It turns out that this eigenvector has approximately equally large components along the remaining fluctuation amplitudes $\delta g_{x,0}$, $\delta\rho_0$ and $\delta\eta_0$. This is consistent with our previous findings, indicating that the density mode, which was alluded to in section \ref{<sec:StabDisorderedState>} and which turns unstable at $\rho=\rho_c$, renders the state of homogeneous collective motion unstable to   fluctuations of the magnitude of the momentum current density. We further note that this picture is in agreement with previous numerical \cite{Gregoire:2004uf} and analytical \cite{Bertin_long} results (cf. \ref{<fig:IlluMotivation>}).

The stability regions of the \emph{grand canonical} model strongly deviate from the above picture. Setting $\psi = 0$ (longitudinal perturbations), we calculated the largest eigenvalue $s_{\mbox{\tiny max}}^{(\mbox{\tiny GC})}$ of the linear system of equations \eqref{<eq:dens_dynamics_GC1>} -- \eqref{<eq:dens_dynamics_GC3>}:
\begin{equation}
\Re\left[s_{\mbox{\tiny max}}^{(\mbox{\tiny GC})}\right] = -\frac{(1-\rho_s)\,q^2}{4\left[(1-\rho_s)^2+\rho_s^2\left(\frac{14}{15}+\frac{2}{3}e^{-2\sigma^2}\right)\right]},
\end{equation}
which is always negative since $\rho_s<1$. In contrast to the \emph{canonical} model, longitudinal perturbations thus always decay exponentially fast in the \emph{grand canonical} model. For perturbations in transverse directions, in contrast,  a positive eigenvalue can be found for sufficiently low noise levels $\sigma<\sigma_r$, with the fastest growing modes posessing wavevectors $\vec q\,\perp\,\vec g_0$.  \ref{<fig:Spatial_analysis_C_Long_GC_trans>}b shows the eigenvalue of the most unstable modes, which occur for $\sigma = 0$. To assess the implications of this instability for the dynamics of the various fluctuation amplitudes, we numerically examined the eigenvector corresponding to the positive eigenvalue, evaluated at the most unstable wavenumber. For densities in the vicinity of the ordering transition, we find that this eigenvector has approximately equal components in both momentum current density fluctuation amplitudes, $\delta g_{x,0}$ and $\delta g_{y,0}$, but an essentially vanishing component along the direction of density fluctuations $\delta\rho_{c}^{0}$. The corresponding instability can thus be classified as a hybrid shear/splay instability, leaving the spatially homogeneously distributed  particle density virtually unaffected. 

\begin{figure}
\centering
\resizebox{0.5\textwidth}{!}{%
  \includegraphics{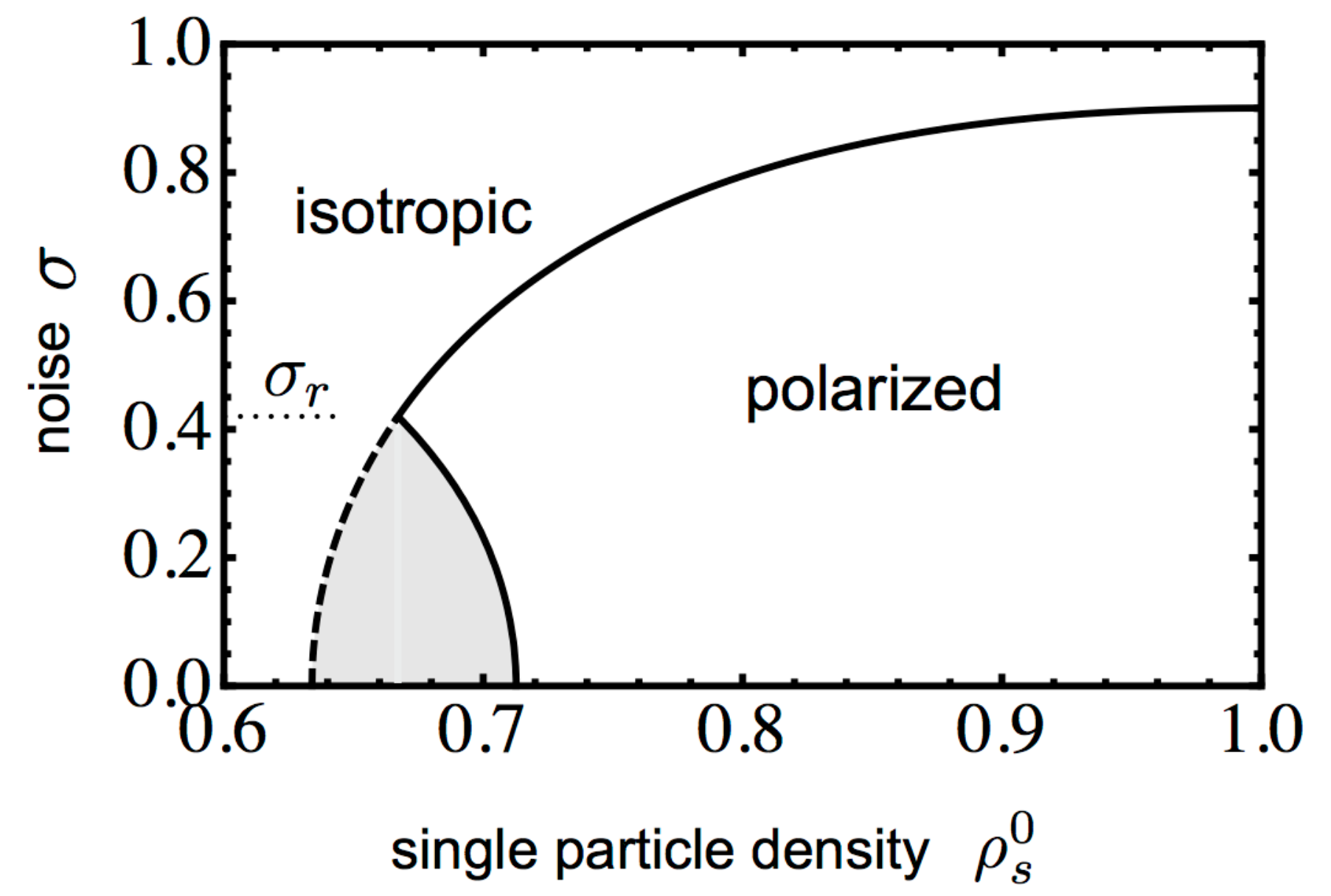}
}
\caption{Phase diagram determined from the homogeneous equations of the \emph{grand canonical} model, as a function of noise level $\sigma$ and single particle density $\rho_s^0$,  now  complemented by the results obtained from the stability analysis of the linearized inhomogeneous equations: Whereas  longitudinal perturbations decay within the homogeneous phase boundary, 
there is a zone (grey shaded) where transversal modes become linearly unstable. 
The width of this zone gradually decreases for increasing noise values $\sigma$, and  vanishes completely above some critical noise value $\sigma_r$ (horizontally dotted line).
}
\label{<fig:Spatial_homogen_with_trans_GC>}       
\end{figure}

Three remarks are in order: First of all, for noise values $\sigma_c(\rho\rightarrow\infty)\!>\!\sigma\!>\!\sigma_r$, the state of homogeneous collective motion becomes linearly stable with respect to arbitrary perturbations, including transverse perturbations. Secondly, even the most unstable eigenvalues ``restabilize'' for densities 
 which are in the vicinity of the ordering transition threshold, which is depicted in  \ref{<fig:Spatial_homogen_with_trans_GC>}.
Finally, a restabilization can also be ``observed'' for the longitudinal instability in the \emph{canonical} model. In this case, however, the restabilization occurs for relatively large densities and thus lies outside the range of validity of the linearized equations \eqref{<eq:canonical_linear1_Inhomogen>} -- \eqref{<eq:canonical_linear4_Inhomogen>}.

\section{Discussion and conclusion}\label{<sect:conclusion>}

To conclude, we discuss and summarize our main findings. To study the onset of collective motion in active media, we started out with a simplified model for a system of self-propelled rod-like particles of variable aspect ratio. Collective motion was assumed to be established in a completely self-organized fashion solely by means of interactions among the constituent particles and in the  absence of any external alignment fields. These interactions were assumed to occur via binary, inelastic particle collisions during which the rods align their direction of motion. Moreover, interactions were assumed to be subject to noise, which we controlled by a single model parameter $\sigma$. To assess some of the structural properties of such systems, we associated each of the particles with one of two classes: \SPS~and \CPS, each with the corresponding density fields denoted as $\rho_s$ and $\rho_c$. 
The class of \CPS~hosts all particles belonging to some coherently moving group of particles, which we referred to as cluster. The rest of the particles can be imagined to make up an isotropic sea of particles and are associated with the class of \SPS. Using this classification scheme, we implemented simple interaction rules, representing cluster nucleation, cluster growth and cluster evaporation; the latter is assumed to occur at some fixed rate $\epsilon$.

To illuminate the self-organization of collective motion, we set up an analytical, kinetic description of such systems, focusing on two archetypical modeling frameworks. Firstly, we considered isolated systems in which the total number of constituent particles is a conserved quantity. This case was referred to as the \emph{canonical} model. Secondly, we examined open systems, which we referred to as the \emph{grand canonical} model. Open systems are in contact with a particle reservoir which keeps the density of \SPS~at a constant level.

Inspecting the corresponding hydrodynamic equations, we were able to establish the following physical picture, portraying the formation of collective motion via dissipative particle interactions: For both, the \emph{canonical} and the \emph{grand canonical} model, we identified two characteristic density scales $\bar\rho$ and $\rho^{(c)}(\sigma)$, with $\rho^{(c)}(\sigma)>\bar\rho$, which allowed us to distinguish three density regimes.

For low densities, $\rho<\bar\rho$, the rate at which particles collide is much smaller than the rate at which clusters disassemble. In terms of a particle based picture, this regime corresponds to a situation, where particle clusters are unstable, evaporating shortly after their nucleation. In the stationary state, the vast majority of particles populates the \SP~phase, rendering the system homogeneous and isotropic even on mesosopic scales. This low density regime terminates at the characteristic density $\bar\rho$, where both classes exchange particles at equal rates.

In the contiguous regime of intermediate densities,  $\bar\rho<\rho<\rho^{(c)}$, the overall rate of cluster formation and growth outstrips the rate at which clusters evaporate, and the majority of particles becomes organized in clusters. Translated to a particle based notion, clusters grow to finite sizes and persist over macroscopic time scales. 
Clusters of coherently moving particles now dominate the physical picture on mesoscopic scales. Yet, interactions among clusters are too rare to establish a macroscopic state of collective motion. On hydrodynamic length scales, the system can be viewed as a homogeneous and isotropic sea of clusters.

For densities exceeding the critical density, $\rho>\rho^{(c)}(\sigma)$, collisions within the cluster phase occur at sufficiently high rates, and macroscopic collective motion emerges. The homogeneous and isotropic state, which has been shown to be stable within the two preceding regimes, thus gets unstable and rotational symmetry is spontaneously broken. Resorting to a particle based image, we can imagine the mean cluster size to reach a ``percolation threshold'', leading to coagulation and net alignment between clusters.

While the qualitative features of the \emph{canonical} and the \emph{grand canonical} model are the same in the low and the intermediate density regime, the establishment of collective motion in the high density regime differs in important respects in both models.
We found that in the \emph{grand canonical} model, a broadly extended region in parameter space exists, where a spatially homogeneous state of macroscopic collective motion exists and is actually stable. Except density, the key parameter controlling the stability of a spatially homogeneous flowing state is the noise amplitude $\sigma$. For low noise levels the homogeneous flowing state gets unstable toward transverse perturbations (i.e. perturbations with wavevectors $\vec q$ perpendicular to the direction of the macroscopic flow). We note, however, that these instabilities are remarkably weak, i.e. the corresponding growth rates are smaller than those of the longitudinal instability by a factor of $\sim10$ (cf. figure \ref{<fig:Spatial_homogen_with_trans_GC>}), and ``restabilization'' of the spatially homogeneous flowing state occurs upon increasing the density only slightly beyond the threshold $\rho^{(c)}(\sigma)$. Interestingly, this transverse instability vanishes altogether, if angular diffusion is slightly enhanced upon increasing $\sigma$. Hence, for intermediate values of $\sigma$, the system directly establishes a homogeneous state of collective motion, which is stable against arbitrary perturbations of small magnitude. Finally, if the noise is too strong, order is destroyed and the system remains isotropic even for arbitrarily large densities. This last statement is, of course, shared among all active systems \cite{Vicsek}, particle conserving or not, and thus applies equally well to the \emph{canonical} model.

In the case of the \emph{canonical} model, a spatially homogeneous base state is unstable toward longitudinal perturbations (i.e. perturbations with wavevectors $\vec q$ parallel to the direction of the macroscopic flow) for all values of the noise parameter $\sigma$. Both, the magnitude of the macroscopic velocity field and the particle density are prone to this kind of instability. This is in agreement with previous analytical \cite{Bertin_long} and numerical \cite{Gregoire:2004uf} results for particle conserving systems, where the emergence of solitary wave structures has been reported in the vicinity of the ordering transition $\rho\gtrsim\rho^{(c)}(\sigma)$. The longitudinal instability thus seems to be a quite generic feature of particle conserving systems with hard core interactions. For an interesting counter example we refer the reader to Ref. \cite{Peshkov:2012uu}, where a particle conserving system with topological interactions has been studied.

We can now combine our findings for both, the \emph{canonical} and the \emph{grand canonical} model, to offer the following mechanistic explanation concerning the emergence of the longitudinal instability. The prerequisite, underlying the establishment of coherent motion, is embodied by two basic processes: Cluster nucleation by collisions among \SPS, and cluster growth by alignment of \SPS~to clusters. Only if, by virtue of these processes, the concentration of \CPS~grows sufficiently large, clusters are able to synchronize their movements by coagulation and macroscopic collective motion emerges. 

Now consider the effect of a density fluctuation in an otherwise homogeneous state of macroscopic collective motion. In the \emph{grand canonical} model, where the density of \SPS~is kept fixed by virtue of a particle reservoir, this fluctuation occurs within the class of \CPS. We can use the right hand side of  equation \eqref{<eq:HydrodynamicEqsHomogeneous2G>}, to assess the implications of such a fluctuation on the local composition of the system in terms of \CPS~and \SPS:  
\begin{equation}
\left(\rho_{s}^{0}+\rho_c\right)\rho_{s}^{0} = \rho_c.
\end{equation}
Note that this equation captures the balance of the two particle currents  between the \SP~and the \CP~phase in the stationary limit. As can be seen from this equation, locally enhancing the density of \CPS~implies a net current from the \CP~phase into the \SP~phase, thus counteracting the effect of the original density fluctuation. Conversely, locally diminishing the density of \CPS~leads to the opposite effect. Density fluctuations are thus damped in the \emph{grand canonical} model and do not impact the macroscopic velocity field, which is set up by the \CPS.

Exactly the opposite happens in the particle conserving \emph{canonical} model. Again, consider a spatially homogeneous base state of macroscopic collective motion. Particles are then distributed among the phases of \CPS~and \SPS~as determined by the balance equation [cf. \eqref{<eq:HydrodynamicEqsHomogeneous2>}]
\begin{equation}
\rho(\rho-\eta) = \rho + \eta,
\end{equation}
where the left hand side describes cluster nucleation and condensation, and the right hand side corresponds to cluster evaporation. This can be seen by using the definitions of the relative density $\eta=\rho_c-\rho_s$, and the total particle density  $\rho=\rho_s+\rho_c$. Now, consider a fluctuation in the total density $\rho$, where, for the sake of simplicity, we assume the relative density $\eta$ to remain constant. In regions, where the fluctuation leads to an increase in the total density by a factor $k>1$ we have
\begin{equation}
k\rho(k\rho-\eta) > k\rho + \eta.
\end{equation}
Hence, the particle current into the \CP~phase grows. As a consequence, the local value of the momentum current density increases, since the \CPS~are the ``carriers'' of the macroscopic momentum. In contrast, in regions, where the fluctuation decreases the total density by a factor $k'<1$ we have
\begin{equation}
k'\rho(k'\rho-\eta) < k'\rho + \eta.
\end{equation}
There the \CP~phase gets depleted and the local magnitude of the momentum current density declines. As a result, high density regions move at faster speeds than low density regions, gathering more and more particles on their way through the system. Conversely, lower density regions continually lose particles to the faster high density structures. In particle conserving systems, every density fluctuation thus automatically triggers a corresponding fluctuation in the momentum current density, which in turn amplifies the density fluctuation. As a result of this process, high density bands of collectively moving \CPS~might emerge \cite{Gregoire:2004uf}. These bands being interspersed by regions where the particle density has fallen below the critical density $\rho^{(c)}$ (and possibly below $\bar\rho$), leading to local destruction of clusters and collective motion.

We close by adding some remarks on the importance of the particles' shape on the establishment of collective motion on hydrodynamic scales. We found that the impact of particle shape on the macroscopic properties of such systems is purely quantitative in the framework of our present study: Varying the particles' aspect ratio results in a shift of the characteristic density scales $\bar\rho$ and $\rho^{(c)}(\sigma)$, which we quantified in equation \eqref{<eq:PvsXi>}. Qualitatively, our conclusions concerning the macroscopic properties of these systems remain unaffected by a change in the particles' aspect ratio. Note that, in our approach, the aspect ratio basically determines the total scattering cross section and thus ``merely'' impacts the rate at which particles collide. We stress, however, that in real systems particle shape is likely to have a profound impact on the entire physical picture of particle interactions, and not just on their rate. The study of those effects lies outside the scope of our present work and would be an interesting topic for future research.

\begin{acknowledgements}
The authors would like to thank Simon Weber and Jonas Denk for discussions and critical reading of our manuscript. Financial support from the DFG in the framework of the SFB 863 and the German Excellence Initiative via the program Nanosystems Initiative Munich (NIM) is gratefully acknowledged. This research was also supported in part by the National Science Foundation under Grant No. NSF PHY11-25915.
\end{acknowledgements}

\newpage
\begin{appendix}
\section{Derivation of Boltzmann collision cylinder for driven rods}\label{<A>}\label{<sect:coll_cylinder>}
\label{<app:stosszylinder>} 

In the framework of our Boltzmann-like description, binary collisions, such as equations \eqref{<SStoCC>}, \eqref{<SCtoCC>} and \eqref{<CCtoCC>}, occur with a certain rate $\Gamma$ depending on particle shape ($L$ and $d$), relative angle of both collision partners $\theta_{12}=|\theta_1-\theta_2|$  and the constant velocity $v$. The quantity  $\Gamma(L,d,\theta_{12})$ characterizes the collision area per unit time -- more commonly referred to as Boltzmann collision cylinder. 
On the scale of the Boltzmann equation, binary collisions occur locally, say in an infinitesimal volume element centered at $\vec r$. Assume that particle $1$ has an orientation $\theta_1$. Then, $\Gamma(L,d,\theta_{12}) \, dt$ gives the area around particle $1$ in which every particle with orientation $\theta_2$ will collide during a time interval $[t,t+dt]$ with particle $1$ . As a consequence  $\Gamma(L,d,\theta_{12}) \, f(\vec{r},\theta_1,t)\,  f(\vec{r},\theta_2,t) d\theta_1 d\theta_2$ equals the number of collisions per unit time and unit area at time $t$, with $f(\vec{r},\theta,t)$ denoting the one-particle distribution function.

\begin{figure}[b]
	\centering
	\resizebox{0.3\textwidth}{!}{%
		 \includegraphics{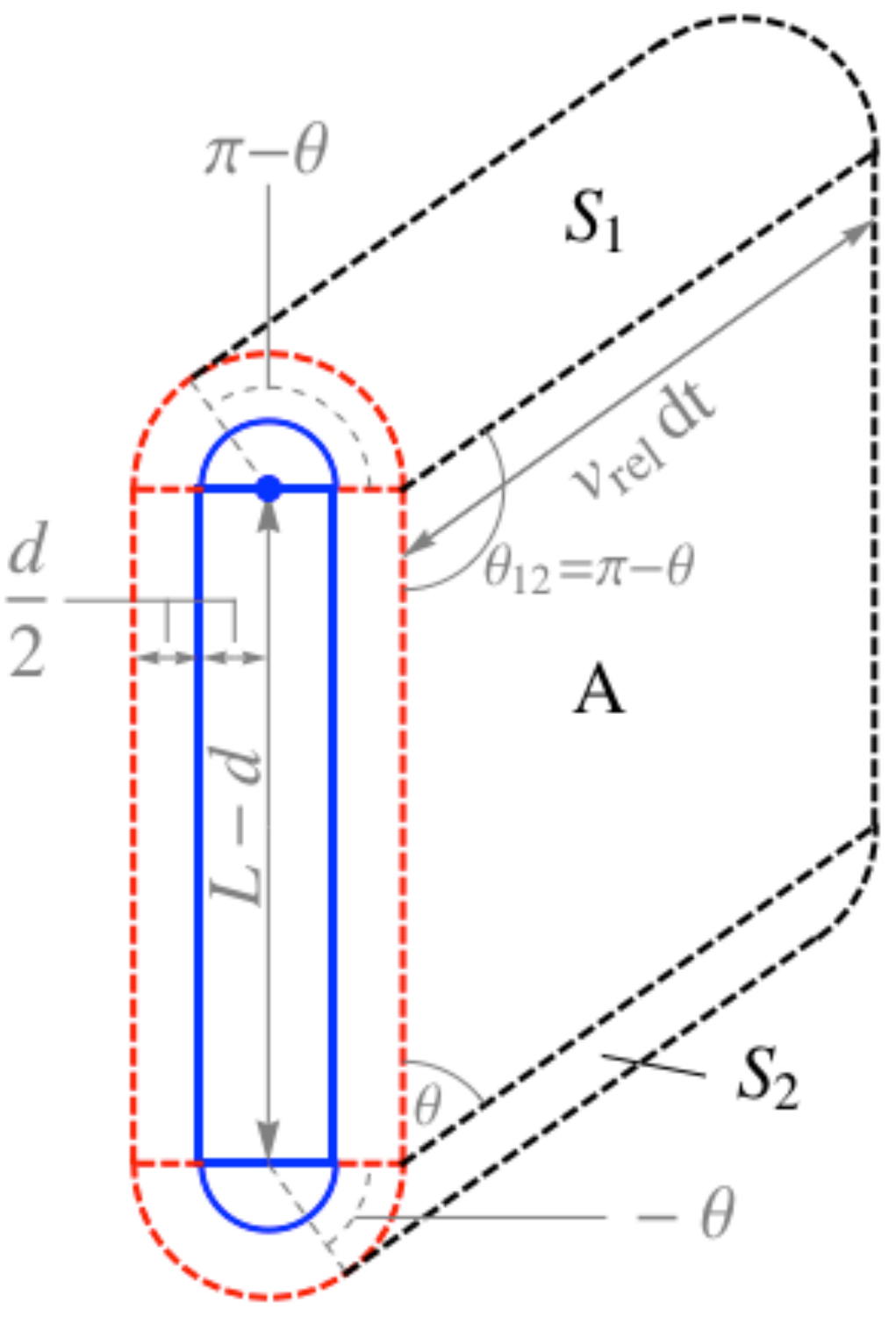}}
	 	\caption{Illustration of the collision cylinder in the rest frame of the blue rod. The red lines indicate the excluded volume due to the finite expansion of the rods. THe quantity $v_{rel}$ denotes the magnitude of the relative velocity of those rods making a relative angle $\theta_{12}=\pi - \theta$ with the blue rod's axis, and is given by $v_{rel}=v|\hat{\vec{e}}(\theta)-\hat{\vec{e}}(0)| =2 v|\sin(\theta_{12}/2)|$.
	}
	\label{<fig:collision_cylinder>}
\end{figure}

To determine $\Gamma(L,d,\theta_{12})$, we take a microscopic point of view. 
Since the model employed in this work assigns to each particle a velocity vector pointing along its rod axis, we can distinguish ``head'' and ``tail''. Referring to  figure \ref{<fig:collision_cylinder>}, without loss of generality we assume $\pi-\theta_{12}\equiv\theta \in [0, \pi]$ (negative relative angles lead to the same result), and consider the blue rod, with the position of its head  indicated by the blue dot. All rods of relative orientation $\theta_{12}=\theta_1-\theta_2$, and with their heads  lying in the area $S=A\cup S_1 \cup S_2$ at time $t$, will collide with the blue rod during the time interval $[t,t+dt]$.
Since $A$, $S_1$ and $S_2$ are disjoint, 
\begin{equation}\label{<eq:three_areas>}
	|S|=|B|+|S_1|+|S_2| \, ,
\end{equation}
where $|X|$ denotes the area of the region $X$. The respective areas are given by:
\begin{equation}
|A| = dt \,v_{rel} \,  (L-d) |\sin{\theta}| = dt \,v_{rel} \,  (L-d) |\sin{\theta_{12}}|,
\end{equation}
and
\begin{equation}
	|S_2|+|S_1| =dt \, v_{rel} \, d \int_{-\theta}^{\pi-\theta}d\phi\,\sin(\phi+\theta)
	=2\, dt \, v_{rel} \, d \, .
\end{equation}
Returning to the laboratory frame we have $v_{rel}=v|\hat{\vec{e}}(\theta_1)-\hat{\vec{e}}(\theta_2)| = 2\,v |\sin(\theta_{12}/2)|$. Noting that $\Gamma= |S|/dt$ (cf. eq. (\ref{<eq:three_areas>})), we find:
\begin{equation}\label{<eq:collision_cylinder_app>}
	\Gamma(L,d,\theta_{12})= 4v\, d \; \left|\sin\left(\frac{\theta_{12}}{2}\right)\right|\left(1+\frac{L/d-1}{2}\left|\sin\theta_{12}\right|\right).
\end{equation}

\begin{figure}
\begin{centering}
\resizebox{0.48\textwidth}{!}{%
  \includegraphics{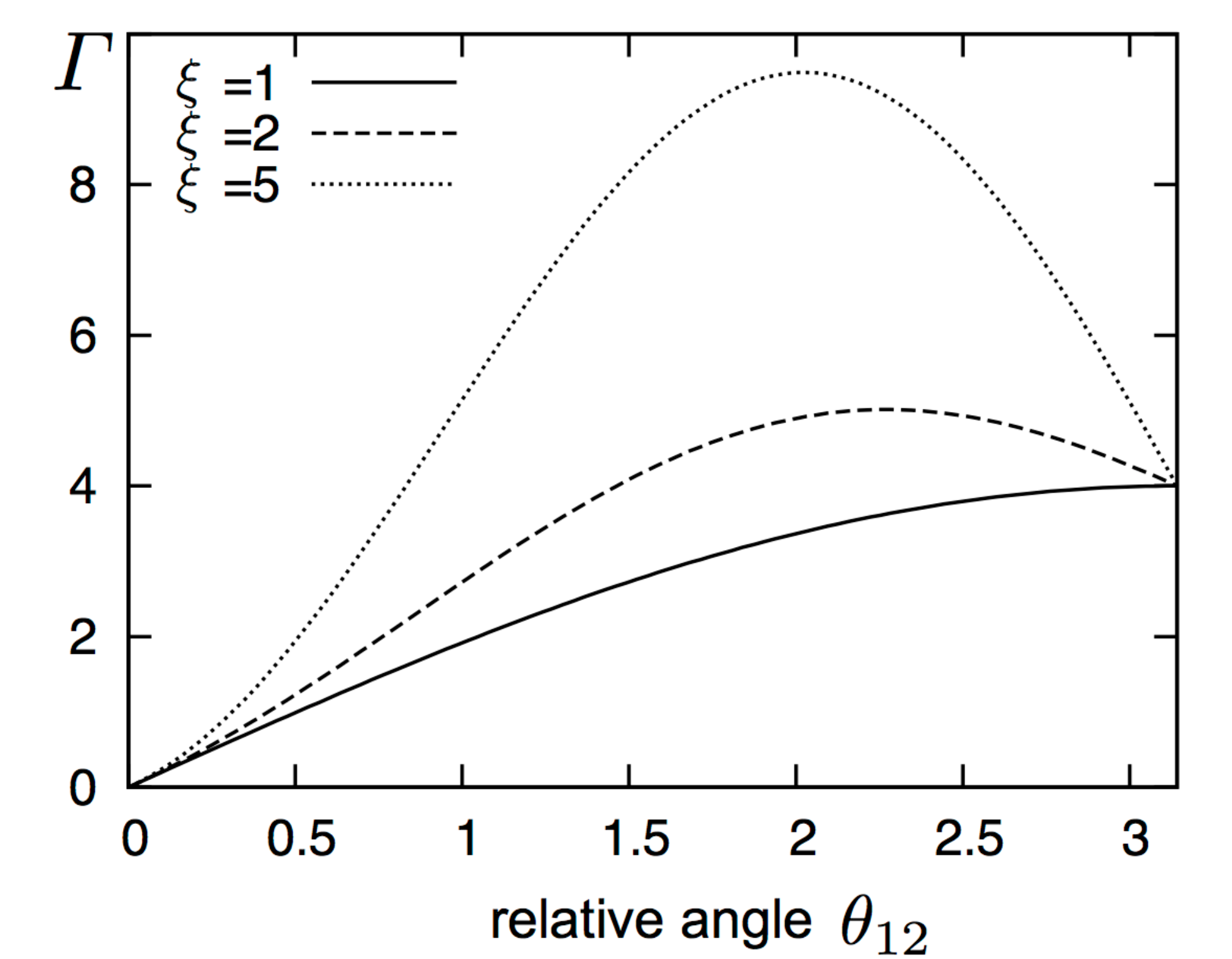}}
\caption{$\Gamma(L,d,\theta_{12})$ as a function of the relative angle $\theta_{12}$ for different values of aspect ratio $\xi$. For the figure, we chose for particle width $d=1$ and for particle velocity $v=1$. Increasing the aspect ratio $L/d$, the most probable collision approaches $\theta_{12}=\pi/2$, whereas for $L/d=1$ the most probable collision is the head-head collision with $\theta_{12}= \pi$.} 
\label{<fig:Boltzmann_surface_xi>}    
\end{centering}   
\end{figure}

In figure \ref{<fig:Boltzmann_surface_xi>}, $\Gamma(L,d,\theta_{12})$ is shown as a function of relative angle $\theta_{12}$ for different particle lengths, whereby the particle width $d$ is kept fixed. Increasing $L/d$ shifts the most probable collision from $\theta_{12}= \pi$ for $L/d=1$ (the case of a sphere; $\theta_{12}=\pi$ leads to the largest value of the relative velocity), towards $\theta_{12}=\pi/2$ for $L/d \to \infty$ (limiting case of a needle; largest target area for $\theta_{12}=\pi/2$).

\section{Derivation of the gradient terms in the hydrodynamic equations}\label{<B>}
To assist the reader in tracing back the emergence of the gradient terms in the hydrodynamic equations \eqref{<eq:HydrodynamicEqsTemp1>} -- \eqref{<eq:HydrodynamicEqsTemp3>} [and, likewise, in equations \eqref{<eq:HydrodynamicEqs1>} -- \eqref{<eq:HydrodynamicEqs3>} and \eqref{<eq:HydrodynamicEqsTemp1G>} -- \eqref{<eq:HydrodynamicEqsTemp3G>}], we briefly summarize the main steps in the derivation of these equations. All gradient terms in the hydrodynamic equations ultimately arise from the convection term in the first line of equation \eqref{<eq:dtc1>} and the closure relation obtained by quasi-statically approximating \eqref{<eq:dtc2>}. Here we collect all such (complex) gradient terms and give a brief derivation of their vector-analytic counterparts. As in the main text, we identify $\mathbb{C}$ and $\mathbb{R}^2$, i.e.
\begin{equation}
\label{eq:translation_c_r2}
f=f_x+if_y\in\mathbb{C} \leftrightarrow \vec{f}=
\begin{pmatrix}
f_x\\
f_y
\end{pmatrix}
\in\mathbb{R}^2.
\end{equation}
To distinguish (genuinely) complex from purely real quantities, we assume $f\in\mathbb C$ and $\rho\in\mathbb R$ in the following.

\subsection*{$(\partial_x+i\partial_y)\rho$}
Using \eqref{eq:translation_c_r2} we immediately obtain
\begin{equation}
\label{eq:gradient}
(\partial_x+i\partial_y)\rho\equiv\nabla\rho.
\end{equation}

\subsection*{$(\partial_x-i\partial_y)(\partial_x+i\partial_y)f$}
By straightforward expansion we find
\begin{equation}
\label{eq:real_laplace}
(\partial_x-i\partial_y)(\partial_x+i\partial_y)f=(\partial_x^2+\partial_y^2)f\equiv\nabla^2\vec f.
\end{equation}

\subsection*{$(\partial_x-i\partial_y)f^2$}
Decomposing $f$ into real and imaginary part and expanding, we find
\begin{eqnarray}
\label{eq:complex_gradient_fsquared}
(\partial_x-i\partial_y)(f_x^2-f_y^2+2if_xf_y)&=&\partial_xf_x^2-\partial_xf_y^2+2\partial_y(f_xf_y)\\
\nonumber
&&+i\left[\partial_yf_y^2-\partial_yf_x^2+2\partial_x(f_xf_y)\right]\\
\nonumber
&\equiv&2\left[\partial_i(f_if_j)-\frac{1}{2}\delta_{ij}\partial_i \vec{f}^2\right]\vec{e}_j\\
\nonumber
&=&2\vec{f}(\nabla\cdot\vec{f})+2(\vec{f}\cdot\nabla)\vec{f}-\nabla\vec{f}^2,
\end{eqnarray}
where $\vec{e}_j$ denotes the $j$-th Cartesian unit vector.

\subsection*{$\left[(\partial_x+i\partial_y)f\right]\left[(\partial_x-i\partial_y)\rho\right]$}
Expanding and collecting real and imaginary parts, we find
\begin{eqnarray*}
\left[(\partial_x+i\partial_y)f\right]\left[(\partial_x-i\partial_y)\rho\right]&=&\partial_xf_x\partial_x\rho-\partial_yf_y\partial_x\rho+\partial_xf_y\partial_y\rho+\partial_yf_x\partial_y\rho\\
&+&i\left(-\partial_xf_x\partial_y\rho+\partial_yf_y\partial_y\rho+\partial_xf_y\partial_x\rho+\partial_yf_x\partial_x\rho\right).
\end{eqnarray*}
Thus,
\begin{eqnarray}
\left[(\partial_x+i\partial_y)f\right]\left[(\partial_x-i\partial_y)\rho\right]&\equiv&\bigl((\partial_if_j)\partial_i\rho+(\partial_jf_i)\partial_i\rho\bigr)\vec{e}_j-(\nabla\cdot\vec{f})\nabla\rho\\
\nonumber
&=&\bigl[(\nabla\vec{f})+(\nabla\vec{f})^t\bigr]\nabla\rho-(\nabla\cdot\vec{f})\nabla\rho.
\end{eqnarray}

\subsection*{$f^2(\partial_x-i\partial_y)\rho$}
We find
\begin{equation*}
f^2(\partial_x-i\partial_y)\rho=f_x^2\partial_x\rho-f_y^2\partial_x\rho+2f_xf_y\partial_y\rho+i\bigl(-f_x^2\partial_y\rho+f_y^2\partial_y\rho+2f_xf_y\partial_x\rho\bigr).
\end{equation*}
Hence,
\begin{equation}
f^2(\partial_x-i\partial_y)\rho\equiv2f_if_j\partial_j\rho\,\vec{e}_i-\vec{f}^2\nabla\rho=2\,\vec{f}(\vec{f}\cdot\nabla\rho)-\vec{f}^2\nabla\rho.
\end{equation}

\end{appendix}

\newpage

\section*{References}


\end{document}